\title{The role of decision confidence in advice-taking and trust formation}

\documentclass[12pt]{article}
\usepackage{graphicx}
\usepackage{authblk}
\usepackage{bm}
\usepackage[hyphens]{url}
\usepackage{booktabs}
\usepackage{amsmath}
\usepackage{amssymb}
\usepackage{amsfonts}
\usepackage{amsfonts}
\usepackage{float}
\usepackage[natbibapa]{apacite}
\usepackage [autostyle]{csquotes}
\MakeOuterQuote{"}

\author[1,2]{Niccol\`{o} Pescetelli}
\author[1]{Nick Yeung}
\affil[1]{Department of Experimental Psychology,
                University of Oxford }
\affil[2]{MIT Media Lab, Cambridge, MA}
\date{\today}

\begin{document}
\maketitle

\begin{abstract}
In a world where ideas flow freely between people across multiple platforms, we often find ourselves relying on others' information without an objective standard to judge whether those opinions are accurate. The present study tests an agreement-in-confidence hypothesis of advice perception, which holds that internal metacognitive evaluations of decision confidence play an important functional role in the perception and use of social information, such as peers' advice. We propose that confidence can be used, computationally, to estimate advisors' trustworthiness and advice reliability. Specifically, these processes are hypothesized to be particularly important in situations where objective feedback is absent or difficult to acquire. Here, we use a judge-advisor system paradigm to precisely manipulate the profiles of virtual advisors whose opinions are provided to participants performing a perceptual decision making task. We find that when advisors' and participants' judgments are independent, people are able to discriminate subtle advice features, like confidence calibration, whether or not objective feedback is available. However, when observers' judgments (and judgment errors) are correlated - as is the case in many social contexts - predictable distortions can be observed between feedback and feedback-free scenarios. A simple model of advice reliability estimation, endowed with metacognitive insight, is able to explain key patterns of results observed in the human data. We use agent-based modeling to explore implications of these individual-level decision strategies for network-level patterns of trust and belief formation.
\end{abstract}

\section{Introduction}
Situations in which feedback on others' performance is not immediately available are abundant in our life. In many contexts, including in education and health, we rely on advice but may not have immediate feedback or other objective standards with which to judge the reliability of that advice. Yet in these contexts we must learn to distinguish good from bad advice and consequently which advisors we should trust. How people do this, and how reliably they do so, are open questions. 

\paragraph{}
Many influential models of trust rely on the trustee's characteristics, like their ability or integrity \citep{Mayer1995}. In some situations, the environment might provide contextual cues regarding such characteristics (e.g., individual, institutional or qualification-based reputation of the advisor). However, here we are interested in whether people can learn advice reliability through experience— even in domains where feedback is sporadic, unreliable or completely absent - and what mechanisms underlie trust formation in these contexts. Specifically, we investigate whether people can detect subtle informational differences among their social partners in the absence of any objective feedback and, if so, whether the trustor's internal metacognitive states play a role in their estimation of advice reliability. 

When available, external (reliable) feedback can guide learning through reinforcement learning mechanisms that have been thoroughly described in previous research \citep{Guggenmos2016, Sutton1998, Behrens2008}. But learning may also be needed when such feedback is not available. Our hypothesis is that this seemingly computationally intractable problem can be solved by using two pieces of evidence that are always available when learning about the reliability of other social actors - namely, the advisor's agreement rate with one's own belief and one's own internal decision confidence: If on a given decision we are certain we are correct, then we can equally be certain that anyone who disagrees with us is wrong, and accordingly down-weight their opinion in the future. If on the contrary we make a choice with less confidence, we should still down-weight disagreeing advice, but now to a lesser extent. In other words, we suggest that people can overcome the absence of objective feedback by exploiting the trial-by-trial co-variation between their own internal decision confidence and the actual state of the environment \citep{Pescetelli2016}. We call this heuristic the agreement-in-confidence heuristic. In contrast with other strategies based on classification variability \citep{Weiss2003}, which require repeated observations, this simple mechanism allows people to estimate an advisor's reliability even in one-shot interactions. This strategy is consistent with previous results showing a decrease of the weight put on advice as a factor of distance from one's own opinion \citep{Yaniv2004}.

\paragraph{}
Confidence can be formally defined as the likelihood of being correct given the evidence and the decision made \citep{Fleming2017}. It has been shown to be an important factor that allows cognitive control \citep{Botvinick2001,Fleming2012} and meta-learning \citep{Flavell1979}. Confidence has also been shown to be involved in group decisions and information integration processes \citep{ Bonaccio2006} and the importance of confidence in human interactions is well established. Much of this literature employs the judge-advisor system (JAS) paradigm \citep{Sniezek1989,Yaniv2000,Bonaccio2006}, in which a participant ("judge") is typically asked to give a first answer to a question and is then presented with the opinions of one or more advisors whose advice can then be used to update the initial judgment. Evidence suggests that confident eyewitness testimony is weighted more by members of a jury \citep{Penrod1995,RoedigerIII2012}, confident people are trusted more and are more influential within groups \citep{Swol2005, Zarnoth1997}, and confidence mediates the effect of power on advice taking \citep{Tost2012}. People have been shown to use confidence to arbitrate disagreement irrespective of true accuracy \citep{Hertz2016, Mahmoodi2015, Penrod1995}. However being confident comes to a cost if confidence is not predictive of accuracy (i.e., if confidence is not \textit{calibrated}). For instance people tend to distrust confident but unreliable testimony \citep{Tenney2007}, at least when feedback is provided or easy to acquire \citep{Sah2013}. 

\paragraph{}
The hypothesis explored here is that the role of confidence in social scenarios goes beyond the simple weighting of diverse opinions, and can in addition be used to approximate useful inferences about others. Two lines of reasoning converge to this conclusion. First, in a probabilistic framework, confidence can be interpreted as the subjective estimated likelihood of having made the correct decision, given the internal sensory states and the decision made \citep{Pouget2016, Meyniel2015a, Aitchison2015, Fleming2017}. Second, in many perceptual tasks used in group decision and advice-taking studies, accuracy correlates within-participant with confidence; that is, more confidently expressed judgments are, on average, more likely to be correct \citep{Henmon1911, Koriat2012}. We suggest that these features make confidence signals extremely valuable when objective feedback is missing because they can be used as a proxy for feedback when the latter is absent. We further speculate that subjective confidence can be used (albeit with inherent noise) to infer the likelihood that a social advisor is correct in a similar way that external feedback is used to track advice reliability using simple reinforcement learning rules \citep{Behrens2008}. 

Studies on the false consensus effect \citep{Dawes1989}, whereby people who endorse an answer believe that others are also more likely to endorse it, have shown that it is indeed rational to use one's own belief as a significant sample of the population and to base predictions about others on such beliefs. More recently, \cite{Bahrami2012b} demonstrated that in the absence of trial-by-trial feedback, participants were able to accrue collective benefit - that is, with group performance exceeding that of the most accurate member's performance - if verbal interaction between participants was allowed. In contrast, providing objective feedback on members' trial accuracy but not allowing for verbal communication did not lead to a collective benefit. This is an interesting finding considering that what should ultimately drive trust in a partner and the decision to follow their judgment is their accuracy rate. Previous studies had demonstrated that conveying information about confidence between participants is crucial for collective benefits in group performance \citep{Bahrami2012,Fusaroli2012,Koriat2012}.

\paragraph{}
To the best of our knowledge, studies have so far investigated a person's confidence either as a signal to others (e.g., confident advisors are trusted more \citep{Penrod1995, Swol2005, Bahrami2010}), or as a signal for one self (e.g., confidence reflects a person's perceived likelihood of being correct \citep{Meyniel2015a, Aitchison2015, Fleming2017, Botvinick2001}). The present work's contribution is in exploring how people's own confidence also affects their learning about others and thus how influential received social information will be. The study investigates learning in feedback-free scenarios, which have received little research attention to date (see \citep{Weiss2003, Rouault2019} for exceptions).

We investigate these questions using a perceptual decision paradigm. These paradigms have become popular in recent group decision studies \citep{Bahrami2010, Bang2017a, Sorkin2001} due to the high degree of control that experimenters have over the information provided to participants and the ability to define an optimal behavior. Our findings show that people are indeed able to learn the reliability of an information source even when deprived of external feedback, and indicate that they do so by relying on internal signals of decision confidence. We further show that this simple mechanism works well when the advisor and advisee's judgments are independent, but shows predictable downsides when information between them is correlated.

\subsection{Overview of the current research}
To investigate the question of whether people use metacognitive information to judge the reliability of a source of advice when other cues are unavailable, we devised a computer-based judge-advisor system task in which participants performed a series of simple perceptual judgments where they first gave their response and associated confidence, then were shown the opinion of different virtual advisors with predetermined informational profiles, and finally were asked to state their (perhaps revised) decision and confidence. Crucially we manipulated across participants the presence of objective feedback on a trial-by-trial basis. The feedback group provides a baseline where we expected participants to accurately judge advisor reliability \citep{Behrens2008}. The behavior of interest is whether, in the absence of feedback, participants are still able to track the reliability of different advisors, and whether they use this reliability estimate to weight proffered advice appropriately. Two measures of trust were recorded to allow for dissociations between implicit and explicit behaviors. First, participants were asked to rate explicitly the reliability of advisors every second block of the experiment (see Supplementary Information for details). Second, the degree to which participants updated their judgments and associated confidence from pre- to post-advice provided an implicit measure of judged reliability, with larger updates indicating greater advice influence \citep{Bonaccio2006}. 

This paradigm builds on traditional advice-taking tasks that have commonly used factual knowledge and forecasting tasks \citep{Bonaccio2006, Soll2011, Tenney2007,Yaniv2004} by allowing direct manipulation of the perceptual evidence available to participants (as it is the case in psychophysics). At the same time, the use of "virtual" rather than real human advisors (cf. \cite{Bahrami2010} and \cite{Zarnoth1997}) allows us strict control over the informational content of the social sources. Bayes theorem can be used to compute the precise information carried by each advisor, as a normative framework within which to test our hypotheses and evaluate participants' performance. 

\paragraph{}
Experiment 1 was designed to test the value of using personal confidence information to learn subtle but important advisors' characteristics, specifically advice accuracy and calibration. When feedback is readily available, both accuracy \citep{Behrens2008} and calibration \citep{Tenney2007} have been shown to be valued advice features affecting both advice trustworthiness and influence. By manipulating across participants the availability of trial-by-trial feedback, we show that people are capable of learning subtle cues in their advisors, and both with and without objective feedback they value accurate advice and calibrated advisors. 

We use a simple Bayesian confidence-updating rule to generate formal predictions of different strategies for estimating advice reliability, when applied to the collected data. When objective feedback is available, the model can use the accuracy history of the advisor to estimate its reliability (\textit{Accuracy} variant). When feedback is absent on the contrary, two alternative rules could in principle be used to overcome the absence of objective feedback, but only one is endowed with access to metacognitive information. Both model variants exploit the association existing between accuracy and agreement among observers \citep{Koriat2012}. Assuming judgments are independent and performance is above chance, agreement rate provides an indication of an advisor's accuracy (e.g., two perfectly accurate observers will agree 100\% of the time on the correct answer). Exploiting this feature, a first model variant (the \textit{Confidence} variant) uses a nuanced strategy that weights agreement by subjective pre-advice confidence. According to this strategy, high confidence agreement is more informative than low confidence agreement. We contrasted the \textit{Confidence} variant with a simpler \textit{Consensus} variant, which does not have access to pre-advice confidence and instead simply estimates the reliability of the advisors according to the rate of agreement with each advisor. We show that these simple algorithms capture patterns of the results seen in the human data.

In Experiment 1, advisors' judgements were independent from the participant's judgments. Experiments 2 and 3 extend Experiment 1 with two related aims. The first one is to distinguish between the \textit{Consensus} and \textit{Confidence} variants. The second is to explore inherent limitations in consensus and confidence-based estimates of reliability. Notice that agreement-based models have an inherent flaw in them: An agent using an agreement-based strategy will systematically overestimate the reliability of advisors who agree with the agent when the agent is wrong, or in other words when agent and advisor share the same biases in their judgments. Systematic deviations will occur whenever two people are basing their judgments on the same, incomplete information, or making a decision with the same inherent biases (cf. \citep{Tversky1974}). We expect these situations to be frequent in many everyday scenarios - e.g., online environments - where people can easily self-assort into information-insulated cliques \citep{Sunstein2001}. Key to both aims is manipulation of the coupling between agreement and accuracy by breaking the independence of initial judgment and advice that is present in Experiment 1. We predict that this will lead to predictable distortions in trust and influence, that differ from feedback-based objective reliability of advice, and that the patterns of trust and influence are best captured by a \textit{Confidence} variant.

In this way, we combine experiments with model-informed analyses. The experiments track patterns of trust emerging in human participants in a task in which advice information and perceptual evidence is carefully manipulated. The models identify the patterns of trust and influence that should be observed if people adopted simple learning strategies based on accuracy estimated from feedback, simple agreement or confidence-weighted agreement. In a final section, we use agent-based modeling to provide a preliminary exploration of the way in which the simple mechanisms of trust formation that we identify empirically might generalize to affect network formation in larger groups of interacting individuals.

\section{Experiment 1}\label{exp1}
Experiment 1 investigates whether people learn about subtle advice reliability cues - specifically, advice accuracy and calibration - in environments that lack objective feedback. Four virtual advisors with differing reliability were designed. They differed in accuracy (proportion correct) and calibration (how expressed confidence scales with probability correct \citep{Fleming2014a}). Participants repeatedly experienced each advisor to give them the opportunity to learn about advisor reliability. Learning was examined in terms of both explicit rating of trust in each advisor, and implicit expression of advice influence, to allow for possible dissociations between explicit and implicit trust. The crucial question of interest was whether these measures of trust and influence would be sensitive to differences in advisor accuracy and calibration even in the absence of objective feedback about the correct answer on each trial. By including a second group of participants, who did have access to trial-by-trial feedback, we could also assess the degree to which the putative agreement-in-confidence heuristic provided an adequate vs. imperfect proxy for objective feedback

\subsection{Method}
\subsubsection{Participants}
Volunteers (N = 46, females = 26, age = 23 $\pm$ 0.45) were recruited in exchange for monetary compensation or course credits. Half of the participants were assigned to the Feedback condition and the other half to the No-Feedback condition. The study was approved by local ethical committee. All participants gave informed consent prior to participation. 

\subsubsection{Paradigm}
\begin{figure}[H]
\centering
  \includegraphics[width=\textwidth]%
    {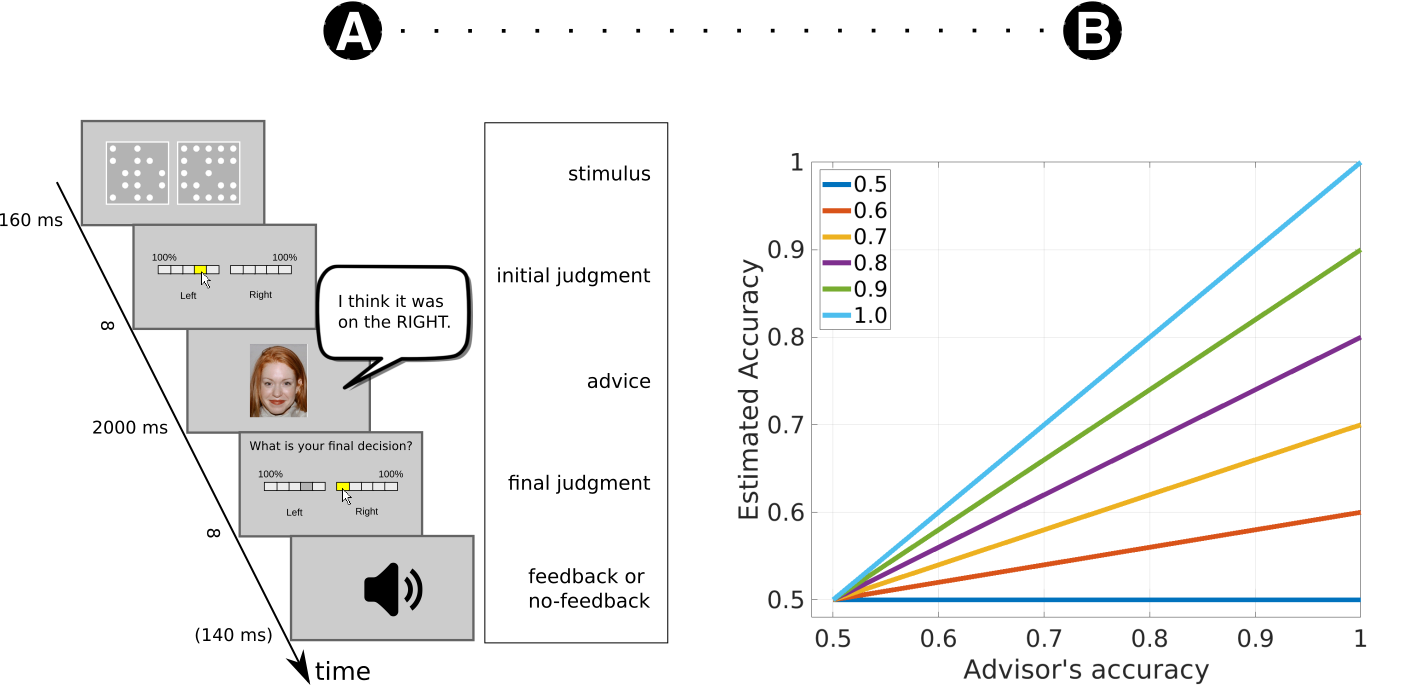}
  \caption[Judge-Advisor paradigm]{\textbf{Panel A}: Experiment 1 paradigm. The task represents a computerised version of the classical judge-advisor system \citep{Sniezek1989}. In our version, the participant first makes a decision about which box contains most dots. The participant expresses his/her opinion on a semi-continuous confidence scale ranging from "100\% sure left" to "100\% sure right". They are then presented with the opinion of a virtual advisor and are asked to update their original judgment with a new rating along the scale. \textbf{Panel B}: A simple model that estimates accuracy from pure agreement rate will understimate the accuracy of any advisor unless the model itself is always correct in its judgments: Estimated accuracy = a*b + (1-a)*(1-b), where a is the objective accuracy of the judge's decisions and b is the advisor's accuracy; i.e., their likelihood of agreeing on the correct answer plus their likelihood of agreeing on an incorrect answer.}
  \label{fig:paradigm}
\end{figure}

The perceptual task used is a dot-count comparison task already described in \cite{Boldt2015}, in which participants are asked to judge which of two boxes contains more dots (Figure \ref{fig:paradigm}). The boxes are presented briefly, for 160 ms, to the left and right of a fixation cross, with dots randomly arranged in a 20 x 20 grid. One box contains more dots, specifically $ndots=200+d$, compared to the other with $ndots=200-d$. By manipulating the $d$ parameter we can control the difficulty of the task. Difficulty was titrated to each individual sensitivity by applying a 2-down-1-up staircase procedure \citep{Treutwein1995} to ensure similar overall accuracy across participants (nominal accuracy rate = 70.7\%). The location of the box with more dots was predetermined in advance by pseudo-randomisation that ensured the number of left and right correct answers was balanced across the experiment. 

After the brief visual presentation of the stimuli (160 ms) participants had unlimited time to enter their response and confidence judgment. They did so in a one-step decision by clicking with the mouse along a semi-continuous scale in 10 steps - ranging from "100\% sure left" to "100\% sure right" - and confirming their response pressing the spacebar. Text landmarks signalling 10\% increases aided the interpretation of the scale. The middle point of the scale (50\% or total uncertainty) was removed and a gap appeared instead, meaning that participants had to commit to one interval (2-alternative forced-choice).

After confirming their response, one out of four different advisors appeared at the center of the screen as a head-shot picture. Pictures were selected from the NimStim database \citep{Tottenham2009} and depicted four Caucasian, smiling female characters that were randomly assigned for each participant to one of the four accuracy/calibration conditions described below. Advice was provided as a pre-recorded female voice through active noise-cancelling headphones. Audio tracks were recorded in from native English speakers and their duration was altered with Audacity\textsuperscript{\textregistered} so that all lasted for exactly two seconds. The association between advisor voice, face and advice profile was randomised across participants to avoid confounds due to appearance or voice characteristics. Advisors could express a binary level of confidence (low vs. high) and either agree or disagree with the participant's judgment. Low confidence was expressed by the sentences "I think it was on the [LEFT/RIGHT]" and "It was on the [LEFT/RIGHT], I think", with one of the two randomly assigned on every trial. Similarly, high confidence was expressed by the sentences "I'm sure it was on the [LEFT/RIGHT]!" and "It was on the [LEFT/RIGHT], I'm sure!". The use of two inverted sentences for each confidence cue ("I'm sure" vs. "I think") was to avoid over-repetition of a single sentence and to balance the differences in emphasis that the English language conveys when using the confidence cues ("I think" vs. "I'm sure") at the beginning or at the end of the sentence. The selection of LEFT or RIGHT depended on the advisor's choice and accuracy as described below. 

After the advice was given, participants had unlimited time to update their decision and confidence level using the same interface and input method as used in the pre-advice period. The question "What is your final decision?" appeared to prompt the update. The pre-advice confidence level remained on the scale as a shaded marker to remind participants of their initial opinion. Participants were allowed to stay in the same position along the scale, increase and decrease their confidence, or even change their minds (i.e., changing interval along the scale). Again they confirmed their final decision by pressing spacebar. In the Feedback condition only, after the final decision was confirmed, a high frequency error tone communicated whether the participant's final decision was incorrect. In the No-Feedback condition, a new trial started immediately after participants had confirmed their final answer. 

At the end of each block, a summary on the post-advice percentage accuracy of the participant (but not of the advisor) was provided to both groups. Notice that advisor presentation was balanced within blocks so the feedback participants received at the end of each block could not favor one advisor over the others. Participants performed 500 trials divided in 10 experimental blocks. Prior to these, two initial blocks with a fifth advisor served as practice and were removed from all the analyses. On each experimental block, each advisor appeared ten times. Ten randomly selected trials within each block were presented with a black silent silhouette and a post-advice decision was not required (null trials). This was done to motivate participants to provide meaningful answers in their pre-advice answers on each trial and avoid pre-advice random guessing. After every two experimental blocks, participants answered a brief questionnaire about their explicit opinions about the four advisers.  Four questions asked participants to directly rate on a scale from 1 (Not at all) to 50 (Extremely) how much they thought each adviser was accurate (Q1), confident (Q2), trustworthy (Q3) and influential on their own choices (Q4) (see SI for complete description).

\subsubsection{Manipulation}
We orthogonally manipulated the average accuracy of the four advisors and their confidence-to-accuracy calibration (Table \ref{table:exp1_paradigm}). We kept the raw numbers of confident and unconfident advice trials equal across advisors (50\% high confidence rate), to avoid people simply trusting the advisor who was the most confident \textit{on average} when objective feedback is not available \citep{Sah2013}. We defined two accurate advisors by setting their accuracy to 80\% and two inaccurate advisors by setting their accuracy to 60\%. Then we set the calibrated advisors to be always accurate whenever confident, and the uncalibrated advisors' confidence to be entirely unpredictive of accuracy. This led to the profiles shown in Table \ref{table:exp1_paradigm}. Calibration is a term used in metacognitive studies indicating the strength of co-variation between confidence judgments and accuracy. Here we quantify calibration as Type 2 $A_{ROC}$ ($A_{ROC}''$), a method which does not make assumptions about the generative model of confidence \citep{Fleming2014a}. Uncalibrated advisors both had an $A_{ROC}''$ of 0.5, meaning that confidence was totally uninformative in predicting the advisor's trial-level accuracy. Due to the experimental design - in which overall accuracy of advisors was fixed at 60\% or 80\%, and calibrated advisors were always correct when high in confidence - calibrated advisors differed in their metacognitive sensitivity according to this metric (Accurate Calibrated advisor $A_{ROC}''$ = 0.72; Inaccurate Calibrated advisor $A_{ROC}'' = 0.84$).

We quantified the informational value of each advisor as the average information gain over possible advice encounters (see Supplementary Information). The mean absolute information gain so computed was lowest for the Inaccurate Uncalibrated advisor, intermediate for the Accurate Calibrated and Accurate but Uncalibrated advisors and the highest for the Calibrated but Inaccurate advisor (Table \ref{table:exp1_paradigm}). This pattern can be intuitively understood by looking at Table \ref{table:exp1_paradigm}: Although the Inaccurate Calibrated advisor's accuracy rate is lower than the Accurate Calibrated advisor, outcomes can be better predicted by her judgment. In particular, her judgments correlate strongly positively with the correct answer when sure and strongly negatively when unsure. On the contrary, when the Accurate Calibrated advisor is unsure there is a much higher uncertainty about the final outcome. A related measure, the expected information gain, which takes into account the frequency of each event (see SI), confirmed a similar pattern of objective advice value.

\begin{table}[h]
\begin{tabular*}{\textwidth}{@{\extracolsep{\fill}}lrrrr@{}} \toprule 
\multicolumn{1}{c}{} & \multicolumn{4}{c}{Advisors} \\ \cmidrule(r){2-5}
Events  & Accurate & Accurate & Inaccurate & Inaccurate \\
count  & Calibrated & Uncalibrated & Calibrated & Uncalibrated \\ \midrule
Incorrect  \\
Confident &  0 & 1 & 0 & 2 \\ \cmidrule(r){1-1}
Incorrect \\
Unconfident & 2 & 1 & 4 & 2 \\ \cmidrule(r){1-1}
Correct  \\
Unconfident & 3 & 4 & 1 & 3 \\ \cmidrule(r){1-1}
Correct  \\
Confident & 5 & 4 & 5 & 3 \\ \midrule
$A_{ROC}''$ & .72 & .5 & 0.84 & .5 \\ 
IG & 0.29 & 0.26 & 0.38 & 0.08 \\ 
IG\textsubscript{e} & 0.063 & 0.063 & 0.084 & 0.021 \\ \bottomrule
\end{tabular*}
\caption[Experiment 1 - Advisors profile]{Experiment 1 - Advisors profile. Values in the central section represent the number of times each event occurred over the course of a 50-trial block (10 null trials, with no advisor, are not shown). Type 2 area under the receiver operating characteristic curve $A_{ROC}''$ represents the metacognitive sensitivity or calibration of each advisor. Information gain $IG$ and expected information gain $IG_e$ quantify the informativeness of the advice received.}
\label{table:exp1_paradigm}
\end{table}

\subsubsection{Exclusion criteria}
An exclusion criterion was set \textit{a priori} for staircase convergence. Participants who showed progressively increasing thresholds (i.e., increasing $d$ values across the experiment) were to be eliminated as this indicated that they were randomly guessing. None of the participants had to be removed when this criterion was applied to our sample. At the end of the experiment the average difficulty parameter $d$ across participants (pooled data) was $9.6 \pm 2.81$. 

\subsubsection{Three models of advisor reliability estimation}
To identify the patterns of trust and influence that would result from adopting simple strategies for advice reliability estimation, we created models that used different types of information to assign value to the advice received. To mimic situations when feedback is provided on a trial-level basis, one model used objective feedback to track each advisor's reliability (the \textit{Accuracy} strategy). Two different strategies can be followed to estimate advisors' reliability when feedback is absent. The first strategy is to use agreement rate with the advisors as an indicator of the advisor's good performance. Assuming that a judge and advisor make independent decisions (below we explore cases where this assumption is violated), then agreement rate will vary as a simple linear function of an advisor's accuracy, with a slope and maximum value that depends on the judge's own accuracy (Figure \ref{fig:paradigm}B). Thus, judges can distinguish between better and worse advisors based on their agreement rates, but will systematically underestimate the accuracy of all advisors to the extent that their own task performance is imperfect.

However, people might solve the task using a more nuanced strategy that leverages the internal sense of confidence as it covaries with objective accuracy \citep{Henmon1911,RoedigerIII2012, Koriat2012}. An effective strategy is to update trust more strongly from judgments when we are certain of our own answer than judgments where we are uncertain or guessing. Suppose, for example, a judge selects the LEFT interval and estimates the probability that they made the correct choice (i.e., their confidence) to be 80\%. Suppose they now receive advice disagreeing with their judgment. Assuming perfect calibration, the judge can then assign a 20\% probability (i.e., $1-0.80=0.20$) that, on that trial, the advisor is correct (i.e., it is 80\% likely that they made an error); and the amount of learning that is possible about the advisors, scales monotonically with confidence. In contrast, a \textit{Consensus} strategy, not using any subjective metacognitive information, would assign to a disagreement trial a 0\% chances that the advisor is correct. Returning to Figure \ref{fig:paradigm}B, a judge learning via confidence-weighted agreement is not confined to estimating advisor reliability according to a fixed linear function that depends on their own accuracy, but rather can learn differentially across trials according to their confidence. An interesting and valuable feature of this approach is that it allows a judge to learn that an advisor is in fact more accurate than they are themselves (something that is impossible with simple agreement-based updating), because good advisors are characterized by agreeing most consistently with the judge in their confidently-made decisions (i.e., those that are most likely to be correct).

Three different model variants (see Supplementary Information for details) account for the Feedback condition (\textit{Accuracy} model), the No-Feedback condition without metacognitive insight (\textit{Consensus} model) and the No-Feedback condition with metacognitive insight (\textit{Confidence} model) respectively. Trial-by-trial agreement, reported confidence and objective feedback from Experiment 1 were used to estimate trust and influence that these model variants would show with each advisor. As is shown below, the three model variants make similar predictions when participants' and advisors' judgments are independent (as in Experiment 1), but generate different results when these initial judgments are correlated (as in Experiment 2-3).

\subsection{Results}
The first set of analyses were performed to show that participants are sensitive to all advice dimensions that were manipulated here, namely accuracy rate, calibration and advice confidence. Of particular interest is whether these main effects vary as a function of feedback availability (i.e., the between-participants manipulation). If the pattern of trust and influence changes as a function of feedback presence vs. absence, this would indicate that different advice dimensions are made more or less salient by the presence of feedback. These questions were investigated through the analysis of both explicit trust ratings and implicit influence measure. Given that in all analyses time was not a significant factor (e.g., when data were pooled across 10 successive blocks to give a factor of time-on-task with 5 levels), the data from all blocks were collapsed together.

\subsubsection{Trust ratings}
\begin{figure}[H]
\centering
  \includegraphics[width=.9\textwidth]%
    {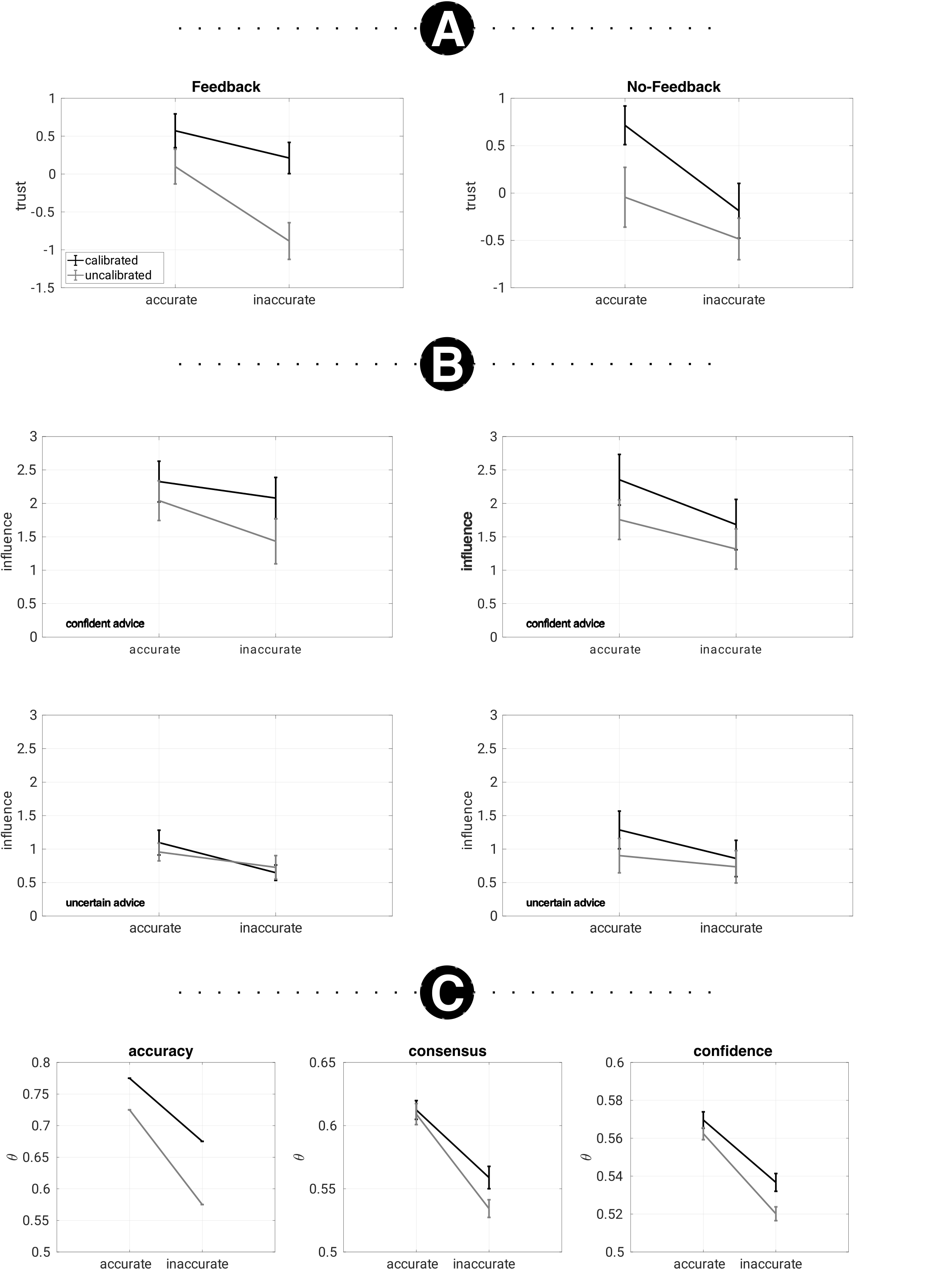}
  \caption[Experiment 1 - influence and trust patterns for humans and simulations.]{Experiment 1 - Trust and influence patterns for humans and simulations. (a) Average trust ratings in the two feedback groups as a function of advisor accuracy and calibration. Error bars represent s.e.m. (b) The effect of advisor Accuracy, advice Confidence and Calibration, and Feedback group on influence measure. Error bars represent s.e.m. (c) Average estimated advisor reliability ($\theta$) according to different variants of the Bayesian model computed from the data of Experiment 1.}
  \label{fig:exp1_all}
\end{figure}

A first analysis focused on participants' explicit ratings of advisor reliability provided at the end of every second experimental block. Ratings were converted into a unitary measure of trust, after preprocessing steps involving normalisation, baseline correction and PCA as described in the Supplementary Information. These steps aimed to remove confounds due to inter-individual differences in responding, to advisor appearance and to question wording. In both groups, questions one and four (regarding advisor accuracy and trustworthiness) scored highest on first PCA component. A mixed-design ANOVA was run to test whether Feedback (between-participants), advisor Accuracy (within-participants) and advisor Calibration (within-participants) affected explicit trust ratings. Results showed significant main effects for advisor Accuracy ($F(1,44)=9.68, p=.003, \eta_G^2=0.079$) and advisor Calibration ($F(1,44)=12.32, p=.001, \eta_G^2=0.076$), but not for Feedback ($F<1$). Participants trusted accurate over inaccurate advice, and calibrated over uncalibrated advice. No interaction term reached significance ($F(1,44)<1.9, p>.16$). Importantly, neither of the main within-participants effects interacted with Feedback, suggesting that participants were sensitive to the reliability of the advice and that this sensitivity did not vary consistently according to the presence or absence of feedback (Figure \ref{fig:exp1_all} A).

We next ran planned 2-way ANOVAs separately for each feedback group, with within-participant factors of advisor Accuracy and Calibration, to assess whether the overall patterns described above were also reliable in each group. In the Feedback group this analysis revealed reliable main effects of both advisor Accuracy ($F(1,22)=8.26, p=.008, \eta_G^2=.09$) and advisor Calibration ($F(1,22)=8.71, p=.007, \eta_G^2=0.12$), but no reliable interaction ($F(1,22)=1.24, p=.27,\eta_G^2=.02$). 

In the No Feedback group, although comparable numerical trends were apparent, neither main effect of Accuracy ($F(1,22)=3.42, p=.07, \eta_G^2=0.06$) and Calibration ($F(1,22)=4.03, p=.05, \eta_G^2=0.04$) reached statistical significance. The interaction term was not significant ($F<1$).

\subsubsection{Influence}
Figure \ref{fig:exp1_all}B plots averaged advisor influence, the implicit measure of learning about advice reliability. Influence was quantified as the signed difference between post and pre-advice confidence (Equation 3, Supplementary Information) and represents the shift in confidence observed after social information. Influence results were analysed using a mixed-design ANOVA that included the same factors of Feedback presence (between-participants), advisor Accuracy and Calibration (both within-participants) as in the trust analysis. For the influence measure an additional within-participants factor was included according to the Confidence expressed by the advisor on a given trial. While both advisor's Accuracy and Calibration are "trait" variables, advice Confidence is a trial-by-trial "state" variable (with values low vs. high). This analysis revealed significant main effects of advisor Accuracy ($F(1,44)=14.80, p<.001, \eta_G^2=0.02$) and Calibration ($F(1,44)=15.84, p<.001, \eta_G^2=0.01$), mirroring the observed effects of advisor reliability seen for explicitly expressed trust. A reliable main effect of advisor Confidence ($F(1,44)=55.82, p<.001, \eta_G^2=.12$) indicated that more confidently expressed advice had greater influence. There was a significant interaction between Calibration and advisor Confidence ($F(1,44)=9.62, p=.003, \eta_G^2=0.004$) indicating that the effect of confidently (vs. unconfidently) expressed advice was greater for calibrated advisors than uncalibrated advisors. The analysis also revealed a significant 3-way interaction between advisor Accuracy, Calibration and Confidence ($F(1,44)=4.75, p=.03, \eta_G^2=8.5e-04$), indicating that the two-way interaction between Calibration and Confidence was larger for inaccurate advisors than accurate ones. Importantly, there was no reliable main effect of Feedback ($F<1$), nor any reliable interaction between Feedback and any other main effects (all $Fs(1,44)<1.1, ps > .29$) suggesting again that participants' sensitivity to advisor reliability, here expressed in terms of the influence of their advice, did not depend significantly on the provision of trial-by-trial feedback.

Although we observed no reliable effects of feedback, we ran planned follow-up ANOVAs for each feedback group separately to assess whether the overall patterns described above were also reliable in each group. In the Feedback group, this analysis revealed significant effects on influence for advisor Accuracy ($F(1,22)=8.23, p=.008, \eta_G^2=0.02$) and Calibration ($F(1,22)=6.35, p=.01, \eta_G^2 = .03$), and advice Confidence ($F(1,22)=32.92, p<.001, \eta_G^2=0.18$). A two-way interaction was found between Calibration and Confidence ($F(1,22)=7.37, p=.01, \eta_G^2=.008$) but not between Accuracy and Calibration nor Accuracy and Confidence (both $F<1$). As shown in Figure \ref{fig:exp1_all}B, the interaction term reflects the fact that the average influence difference between calibrated and uncalibrated advisors was mainly shown for highly confident advice compared to uncertain advice. Finally a reliable three-way interaction between Accuracy, Calibration and Confidence ($F(1,22)=8.32, p=.008, \eta_G^2=.003$) suggests that such interaction was stronger for inaccurate advisors than accurate ones (Figure \ref{fig:exp1_all}B). 
    
A corresponding ANOVA on influence measures in the No-Feedback group revealed main effects of advisor Accuracy ($F(1,22)=6.87, p=.01, \eta_G^2=.02$) and Calibration ($F(1,22)=9.48, p=.005, \eta_G^2=.01$), and advice Confidence ($F(1,22)=22.95, p<.001, \eta_G^2=.07$). The results show that participants were more influenced by Accurate and Calibrated advisors, and by Confident advice compared to uncertain advice, paralleling the effects observed when feedback was provided. None of the two-way interaction terms reached significance (Accuracy rate x Calibration: $F(1,22)=1.35, p>.25, \eta_G^2=.001$; Accuracy rate x advice Confidence: $F(1,22)=2.90, p=.1, \eta_G^2=.002$; Calibration x advice Confidence: $F(1,22)=2.59, p=.12, \eta_G^2=.001$), although numerically there were greater effects of confidence for calibrated advisors (as observed in the Feedback condition). The three-way interaction between Accuracy, Calibration and Confidence was not significant ($F<1$). 

\subsubsection{Model results}
The three model variants were run based on the actual series of each participant’s decisions and (where appropriate) their associated confidence, and the advice they received and (where appropriate) its accuracy, in Experiment 1. We looked at how the different variants of the model fared in evaluating different advisors' profiles based on this input. The aim of the following analyses was to verify how the pattern of final model's trust ($\theta$) in each advisor differed when different pieces of information were used to compute it. The model is not intended as a mechanistic description of participants' behaviour. Instead, it aims to show that a simple reliability estimation rule will develop differentiated patterns of trust across advisors depending on what types of information it relies upon. For this analysis, data from the Feedback and No-Feedback groups were pooled together to increase statistical power, as the presence of feedback did not affect the variables that model's variants were based on, namely advisors' accuracy, agreement rates and participant's pre-advice confidence ratings respectively.

The model's estimates of advisor reliability ($\theta$ parameter) can be used to predict nominal confidence changes using Bayes rule (see Supplementary Information). Confidence changes predicted by the model could have been used to plot the model's implicit trust in the advisors in a manner similar to what already shown with the behavioral data. However, contrary to $\theta$ values, confidence changes are an implicit measure of trust and they are affected not only by the value assigned to an advisor but also by the pre-advice confidence of that trial. Thus, although the two measures are related, $\theta$ represents a direct measure of the model's belief about advisors' reliability (a trial-by-trial measure that is not available in the human data).

\paragraph{}
Each model variant's final $\theta$ values - i.e., the model's advisor reliability estimates on the last trial of the experiment - were analyzed using a 2x2 repeated measures ANOVA with factors of advisor Accuracy (high vs low) and Calibration (high vs. low). Scaling factors were applied to agreement to take into account the fact that in this experiment advisors provided a confidence judgment with their advice. The \textit{Accuracy} variant is, in this experiment, fully pre-determined by the advisors' set accuracy rates and thus no statistical analysis were run due to the absence of variability across participants. We plot however its trust value $\theta$ for visual reference as it shows that when the model is provided with information about the objective performance of the participant (and thus of the advisors), it is able to distinguish advisors both in terms of their Accuracy rate and their confidence Calibration. The difference between calibrated and uncalibrated advisors is larger for inaccurate than accurate advisors, due to the weights used to convey advice confidence (Figure \ref{fig:exp1_all}C). In the absence of externally given objective feedback, both the \textit{Consensus} variant - which estimates advisors' reliability by assuming that advisors are correct whenever they agree with the participant’s own first decision, and wrong otherwise - and the \textit{Confidence} variant - which uses agreement as a proxy for feedback like the \textit{Consensus} variant, but scales them by pre-advice confidence - show greater trust for Accurate ($F(1,45)>165.84, p<.001, \eta_G^2=.44$) and Calibrated ($F(1,45)=32.70, p<.001, \eta_G^2=.13$) advisors compared to inaccurate or uncalibrated ones. Neither variant showed a significant interaction between the two factors ($F(1,45)<2.65, p>.11, \eta_G^2=.01$). Notice that advisors were not constrained to agree with the participant a pre-determined amount of times, thus explaining the variability observed across participants according to the specific sequence of decisions they made and advice they received.

Taken together, these modeling results show that simple computations of advisor reliability perform well at this task even when trial-level feedback is absent, effectively capturing key patterns of trust observed in the human data across feedback conditions. For the task used in Experiment 1, the three variants do not make contradictory predictions on which advisors should be trusted. In particular the two No-Feedback variants (that base trust on simple agreement, or agreement weighted by confidence) cannot be disentangled using the data collected from this experiment, with both showing sensitivity to both accuracy and calibration of an advisor.

\subsection{Discussion}
In this experiment we tested whether participants would distinguish among advisors with differing profiles, and whether patterns of trust and influence would vary according to the availability of objective feedback. Across conditions, participants expressed greater trust in, and were more influenced by, advice characterised by high confidence, high accuracy and high calibration. Although the first two dimensions are not a surprise \citep{Sniezek2001}, the fact that participants also value advice calibration is less well documented. \cite{Tenney2007} showed, in mock jury decisions, that confident but inaccurate witnesses are discredited more than equally inaccurate but less confident ones. \cite{Sah2013} further confirmed that when feedback is readily available, confident but incorrect advice backfires on an advisor's reputation. However, when feedback was not available the same authors found evidence for a confidence heuristic whereby confident advisors were more influential regardless of accuracy \citep{Price2004}. 

The current study extends this previous work in two respects. First, the behavioral as well as the simulation findings highlight the value in using information regarding agreement and confidence in social learning: Although the patterns of trust and influence were not identical across Feedback and No-Feedback conditions, no consistent (statistically reliable) differences were observed, and overall the results indicate that people can learn useful distinguishing characteristics (accuracy and calibration) even in the absence of objective feedback, and even when average advisor confidence is kept constant \citep{Price2004,Sah2013}. Second, we replicate in a highly controlled perceptual task setting the finding that judges value advisor's confidence calibration. By allowing repeated interactions with the same advisors and by fixing advisors' average expressed confidence, the current paradigm was able to quantify advice calibration and show that people value it over and above the overall level of confidence with which advice is expressed. 

\paragraph{}
Simulation results based on straightforward implementations of learning strategies - estimating advice reliability according to objective feedback, agreement, or agreement weighted by confidence - were consistent with the empirical results in showing similar patterns of trust across advisors regardless of the presence or absence of objective feedback. Thus, calculating agreement rates and weighting agreement by subjective confidence can be effective heuristics for estimating advisor reliability when objective feedback is absent. The reason for this is that both agreement and subjective confidence covary with accuracy. Thus, both are useful cues, and in the present data the two are difficult to disentangle. To disambiguate predictions of the two hypotheses, and to explore crucial limitations in the use of agreement and confidence in estimating advisor reliability, the following two experiments included advice that was not independent of participants' initial decisions. Instead, advice varied systematically to allow orthogonal manipulation of (1) advisors' agreement and accuracy rates and (2) advisors' agreement rates and subjective confidence. While Experiment 1 showed the normative value of using such heuristics when advice is independent from a judge's initial decision, in these experiments we explored the impact of these heuristics when advice is not independent.

\section{Experiment 2}\label{exp2}
As shown by the \textit{Consensus} model, the frequency of agreement with the advice might be used as a cue to the reliability of the advice, if observations from individual observers are independent. This observation echoes the Wisdom of Crowds (WoC) effect \citep{Galton1907,Surowiecki2004}, whereby aggregation of individual opinions results in better-than-expert performance. Wisdom of crowds is often cited as the archetype of collective intelligence \citep{Krause2010}. The classic explanation of the effect is that individual judgments are the sum of a signal component and random noise. Averaging individual judgments has the effect of enhancing signal by averaging out noise, with a similar mechanism as repeated measures in statistics \citep{Armstrong2001}.

However, people's judgments are rarely independent and distorted only by random noise \citep{Koriat2012,Krause2010}: We watch the same news, browse the same websites and are affected by the same cognitive biases as others we interact with. Thus, there may be dependencies between judgments. One source of dependence between opinions is the use of similar cognitive heuristics \citep{Tversky1974}. Other sources can arise from using similar strategies to sample information from the environment \citep{Vandormael2017}, being exposed to similar signals \citep{Kao2014, Kao2014a} or belonging to the same network clique \citep{Sunstein2001,Jasny2015,Jamieson2008}. Crowds are known to be susceptible to large error cascades \citep{LeBon1895,Mackay1841}, economic bubbles \citep{DeMartino2013a}, polarisation \citep{Myers1976} and \textit{groupthink} \citep{Janis1972, Turner1998}. An open question is whether these examples of error magnification are due to a lack of judgment independence. Dependencies in error sources may lead to greater consensus among observers but to a decorrelation from actual accuracy - e.g., by agreeing on errors \citep{Koriat2012}. In the context of advice, the focus of the present research, opinions and suggestions might be framed in relation to the person we are advising whether accidentally or deliberately. This framing will influence the extent to which opinions agree or disagree with each other, and thus the information carried by other people's consensus. As a first exploration of biased advice scenarios, we thus investigated whether in these contexts people trust accurate advisors or advisors who tend to agree. 

\paragraph{}
Specifically, advisor accuracy was manipulated so that two advisors were on average highly accurate (around 80\% accuracy) and two advisors were on average relatively inaccurate (around 60\% accuracy). Orthogonally, advisor agreement rate with the participant's judgment was manipulated so as to create two advisors who agreed with the participant frequently (around 80\% of trials) and two advisors who tended to have a lower agreement rate with the participant (around 60\%). This created an advisor who is highly accurate and tends to agree with the participant, an advisor who is highly accurate but tends to disagree with the participant, an advisor who is not very accurate but still tends to agree with the participant, and an advisor who is neither accurate or agreeing. The two intermediate profiles are an important test case where the two dimensions of accuracy and agreement work against each other. The agreeing but inaccurate advisor represent someone who shares biases with the participant and so makes similar (correlated) mistakes. The accurate but disagreeing advisor instead represents an advisor who uses different information so is accurate and tends to be correct when the participant makes mistakes and vice versa. Knowing what advice characteristics are made salient by the presence of feedback can shed light on the mechanisms of trust formation and advice influence.

\subsection{Methods}
\subsubsection{Participants}
The experiment included 46 participants equally divided between the two feedback groups (37 females in total, 18 of whom were in the Feedback group, mean age: $21.63 \pm 3.02$). Participants were recruited through Oxford University's online recruitment platform and local news advertisement. Participants were compensated either with money (10\pounds /hour) or course credits. All participants provided informed consent before the experiment. All procedures were approved by the local ethical committee. 

\subsubsection{Paradigm}
The overall design was very similar to Experiment 1, with advice provided and its impact measured in the context of the same dot-count perceptual decision task. Participants performed ten blocks of 44 trials each. In this experiment, advice was presented in the form of a binary judgment, expressed by the sentences "I think it was on the [LEFT/RIGHT]" and "It was on the [LEFT/RIGHT], I think", depending on advisor's opinion and random selection of one of the two sentences across trials (i.e., there was no across-trial manipulation of advisor confidence). For participants in the Feedback group only, after the final choice was confirmed, a high frequency beep sound (duration 140 ms) was played whenever the participant's final choice was incorrect. 

At the end of every second block, a computer-based questionnaire was presented to participants prompting them to evaluate the four advisors on a 50-point scale (1="Not at all", 50="Extremely"). The questions were the same as those used in Experiment 1 apart from question two (question about advisor's confidence) that was replaced with a question about advisor likeability, to account for the advisors not expressing confidence along with their judgments in this experiment.

In each block, the presentation of the four advisors was randomly shuffled across trials, with each advisor appearing exactly ten times. Four additional trials were randomly inserted into the block and served as null trials, where a silent silhouette was displayed instead, to encourage participants to register meaningful initial confidence judgments as well as final (post-advice) judgments. The difficulty of the first-order task was again titrated to a target level of 70.7\% accuracy using a 2-down-1-up staircase method. Prior to commencing the 10 experimental blocks, participants completed two practice blocks, with ten practice trials each, with a fifth advisor. Practice trials were removed from all the analyses.

\subsubsection{Manipulation}
To disentangle agreement rate from the accuracy of the advisors, the probability of agreement conditional on the participant's choice accuracy was manipulated. Through the staircase procedure it was expected that all participants would converge to an accuracy level of about 70\%. This enabled us to manipulate advisor accuracy and agreement rate by pre-determining the probability of agreement after a participant's error or a participant's correct response. Both accuracy and agreement were manipulated to have two levels (high=80\% and low=60\%). This gave rise to the four advisor profiles defined in Table \ref{table:table_exp2}. Probabilities are expressed as a fraction over the number of participants' expected correct (7) and incorrect (3) judgments, over the number of encounters with one advisor during one block (10).

\begin{table}[H]
\begin{tabular*}{\textwidth}{@{\extracolsep{\fill}}lrrrr@{}} \toprule 
\multicolumn{1}{c}{} & \multicolumn{4}{c}{Advisors} \\ \cmidrule(r){2-5}
 & High Accuracy & High Acc. & Low Acc. & Low Acc. \\
 & High Agreement & Low Agr. & High Agr. & Low Agr.  \\ \midrule
$p(Agr|Cor_s)$  & 6.5/7 & 5.5/7 & 5.5/7 & 4.5/7 \\ \cmidrule(r){1-1}
$p(Agr|Inc_s)$  & 1.5/3 & 0.5/3 & 2.5/3 & 1.5/3 \\ \midrule
Expected \\
Acc. rate & 80\% & 80\%& 60\%& 60\%\\ \cmidrule(r){1-1}
Expected \\
Agr. rate & 80\% & 60\% & 80\% & 60\% \\ \midrule
$IG$ & 0.28 & 0.27& 0.03 &0.06 \\ 
$IG_e$ & 0.09& 0.13 & 0.01 & 0.03 \\ \bottomrule
\end{tabular*}
\caption[Experiment 2 - Advisors' profiles]{Experiment 2 advisors' profiles. Expected accuracy and agreement rates of different advisors are disentangled by manipulating the probability of the advice agreeing with the participant, conditional on the participant's accuracy. In the table, probabilities are expressed as a fraction of the number of participants' expected correct (7) and incorrect (3) judgments, during the number of encounters with one advisor (10) in a single experimental block. Information gain and expected information gain - $IG$ and $IG_e$ respectively - indicate average informational value of the advice, computed as information gain and expected information gain respectively (see Supplementary Information).}
\label{table:table_exp2}
\end{table}

The dissociation between accuracy and agreement was possible by making the \textit{accurate but disagreeing} advisor disagree more often on trials when the participant had made an incorrect initial decision (meaning a correct response for the advisor). Similarly, an \textit{inaccurate but agreeing} advisor was created by making the advisor more likely to agree when the participant's initial judgment was incorrect. For the two advisors for whom accuracy and agreement work in the same direction - i.e., \textit{accurate and agreeing} vs. \textit{inaccurate and disagreeing} - it can be expected that participants will favor the former over the latter. However when accuracy and agreement pattern work in opposite directions - as is the case for the agreeing but inaccurate and the disagreeing but accurate advisors - it can be expected that the availability of external feedback will favor the accurate advisor in the Feedback group but not in the No-Feedback group.

Similar to Experiment 1, we used conditional probabilities and the participants' expected accuracy to compute the informational value of each advisor (Table \ref{table:table_exp2}). Advisors' mean absolute information gain $IG$ and expected information gain $IG_e$ were computed as in the previous experiment. Contrary to Experiment 1, however, advisors did not express different levels of confidence. This created only two possible situations on each trial, either agreement or disagreement. As can be seen in Table \ref{table:table_exp2}, information gain was highest for the accurate advisors and the lowest for the \textit{inaccurate but agreeing} advisor. This advisor is in fact counter-intuitively anti-predictive of the correct answer: In contrast with the other advisors, this advisor's probability of agreement is higher when the participant's initial judgment is incorrect than when it is correct. Thus, an optimal observer would decrease their confidence when in agreement and increase it when in disagreement with this advisor. For all other advisors, agreement should increase confidence and disagreement should decrease it. 

\paragraph{}
The same two measures of interest as in Experiment 1 - explicit trust and implicit influence - were calculated. Similarly, the three model variants were run on the new behavioral dataset. For pairwise comparisons, both p-values and Bayes factors are reported. Bayes factors are reported using the notation suggested in \cite{Dienes2017}. The priors on the effect sizes to compute Bayes factors are informed by results of Experiment 1 (while no priors were available for Experiment 1).

\subsubsection{Exclusion criteria}
The first two experimental blocks were removed from the analysis to allow the staircase procedure to fully adapt to each individual's threshold. This was necessary given that our manipulation was heavily dependent on the expected accuracy rate of the participants. A further exclusion criterion was set to exclude all participants whose threshold never converged, which suggests a random response strategy. None of the participants had to be removed on the basis of this criterion. The perceptual task difficulty $d$ (dot difference between boxes) after staircasing was $9.93 \pm 2.96$ (pooled data).

\subsection{Results}
A first set of analyses explored the effect of our manipulation on the trust and influence measures as a function of whether objective trial-level feedback was provided. In particular we are interested in whether any of the manipulated within-participants factors (advice agreement rate and accuracy) interacted with feedback presence. 

\subsubsection{Trust ratings}
\begin{figure}[H]
\centering
  \includegraphics[width=1\textwidth]%
    {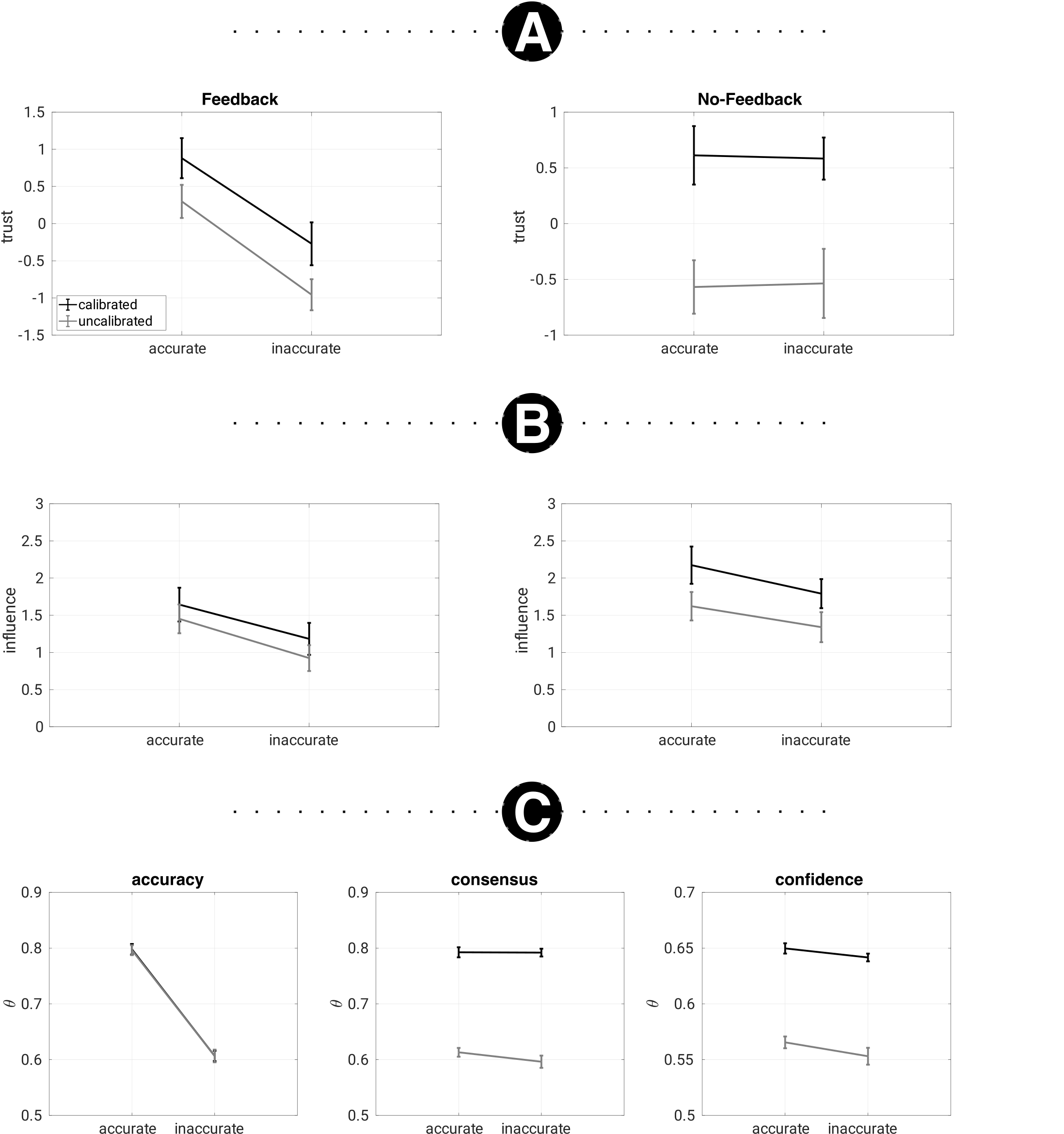}
  \caption[Experiment 2 - Trust ratings and influence of human and simulations.]{Experiment 2 - Trust ratings and influence for human participants and simulations. (a) Average trust ratings in the two feedback groups, divided by advisor accuracy rate and agreement rate. Error bars represent s.e.m. (b) Influence that advice had on participants' opinions in the two feedback groups and divided by advisor accuracy rate and agreement rate. Error bars represent s.e.m.(c)}
  \label{fig:exp2_all}
\end{figure}

A mixed-design ANOVA was run on explicit trust ratings with feedback group as a between-participants factor and advisor accuracy (low vs. high) and agreement rate (low vs. high) as within-participants factors. This analysis revealed significant main effects for both accuracy ($F(1,44)=8.36, p=.005, \eta_G^2=.06$) and agreement rates ($F(1,44)=22.52, p<.001, \eta_G^2=.1$), but not for feedback ($F<1$). A significant interaction was found between feedback and accuracy ($F(1,44)=8.41, p=.005, \eta_G^2=.06$) but not between feedback and agreement rate ($F(1,44)=1.88, p=.17, \eta_G^2=.01$) or between agreement and accuracy ($F<1$), nor a significant three-way interaction ($F<1$). The results suggest that in their explicit ratings of trust, participants reported on average to prefer advice coming from accurate sources and sources that tended to agree with their own judgments. Importantly, however, the effect of accuracy was modulated by the presence of feedback, with effects of advisor accuracy apparent only in the Feedback condition (Figure \ref{fig:exp2_all}A). 

We then looked at trust ratings in the two feedback conditions individually. A repeated measures ANOVA on trust ratings from the Feedback group revealed significant effects of both advisor's Accuracy ($F(1,22)=21.36, p<.001, \eta_G^2=.20$) and Agreement rate ($F(1,22)=5.62, p=.02, \eta_G^2=.06$), but no significant interaction ($F<1$). The effect of accuracy was much stronger than the one observed for agreement as indicated by the generalised eta squared values \citep{Bakeman2005}. Nonetheless, it is notable that agreement had a significant effect on trust even when controlling for accuracy.

A corresponding ANOVA on trust rating data from the No-Feedback group found that Agreement rate ($F(1,22)=18.91, p<.001, \eta_G^2=.18$) but not Accuracy ($F<1$) affected participants' explicit reliability judgments. No significant interaction was found between the two factors ($F<1$). These findings suggest that when feedback was not directly available to estimate partners' reliability, participants expressed greater trust in agreeing advisors over disagreeing ones, but not in accurate advisors over inaccurate ones. 
    
\subsubsection{Influence}
The next analysis focused on influence, as an implicit measure of trust. A mixed-design ANOVA on the influence measure was run with feedback as a between-participants factor and advisor Accuracy (low vs. high) and Agreement rate (low vs. high) as within-participants factors. This analysis revealed significant main effects of advisor Accuracy ($F(1,44)=14.79, p<.001, \eta_G^2=.04$) and Agreement rate ($F(1,44)=13.91,p<.001, \eta_G^2=.03$), and a non-significant interaction between these factors with a numerical trend for participants to be more influenced when feedback was not available ($F(1,44)=3.19, p=.08, \eta_G^2=.04$)(Figure \ref{fig:exp2_all}B). Importantly, no significant interactions were found between Accuracy and Feedback ($F<1$), between Agreement and Feedback ($F(1,44)=2.01, p=.16, \eta_G^2=.005$) and between Accuracy and Agreement rate ($F<1$). No significant three-way interaction was found either ($F<1$). The results show that participants were more influenced by accurate advisors and by advisors characterized by high agreement rates with their own judgments. Importantly, however, the effects did not interact reliably with feedback, suggesting that similar effects were found in the two condition groups independently of feedback availability. We then looked at individual pattern of results within each experimental group.

A two-way ANOVA on influence measures in the Feedback group showed significant effects for both advisor's Accuracy ($F(1,22)=12.71, p=.001, \eta_G^2=.06$) and Agreement rate ($F(1,22)=5.81, p=.02, \eta_G^2=.01$). No significant interaction was observed between the two factors ($F<1$). As expected, when looking at opinion change, participants' opinions were more affected by accurate advisors over inaccurate ones. Interestingly, mirroring effects observed in explicitly rated trust, participants were affected by advisors who tended to agree with their own judgment, although feedback indicated equivalent accuracy rates for pairs of advisors characterized by different agreement rates.
    
When looking at the No-Feedback condition, a similar pattern emerged. A 2x2 repeated measures ANOVA on influence showed significant main effects for both advisor's Agreement rates ($F(1,22)=8.60, p=.007, \eta_G^2=.06$) and Accuracy ($F(1,22)=4.09, p=.05, \eta_G^2=.02$), although the effect size of Agreement was the greater of the two. No significant interaction was observed ($F<1$). Thus, in the No-Feedback group, the results suggest a dissociation between what people reported in the questionnaires, with higher trust reported for agreeing advisors regardless of accuracy, and people's performance in the task, where influence showed effects of both accuracy and agreement rate.

\subsubsection{Model results}
The models described above were applied to data from Experiment 2, to understand whether the \textit{Consensus} and \textit{Confidence} model variants behaved differently in scenarios where advice accuracy and advice agreement rate are dissociated. Each model variant was used to provide an estimate of trust in each advisor based on each participant's expressed confidence, experienced advice and, in the case of the \textit{Accuracy} model, advice objective accuracy. In this experiment, advisors did not express a confidence judgment about their opinions. Thus all model variants could be simplified by not taking into account advice confidence (see Supplementary Information for details). 

Given that we are not trying to fit the model to the human data, but only observing how trust patterns change as a function of the information fed to the model, data from the two feedback groups were pooled together. Trial-by-trial pre-advice confidence and advice were input to each of the three model variants and resulting $\theta$-values for each advisor were compared (Figure \ref{fig:exp2_all}C). Both the \textit{Accuracy} and the \textit{Confidence} models' $\theta$ values show a significant effect of Accuracy ($F(1,45)>8.85, p<.005, \eta_G^2>.05$), while the \textit{Consensus} model only showed a non significant marginal effect ($F(1,45)=3.22, p=.07, eta_G^2=.02$). Both the \textit{Confidence} and \textit{Consensus} models show a significant effect of Agreement ($F(1,45)>434.7, p<.001, \eta_G^2>.71$), but no reliable interaction between the two factors ($F(1,45)<2.04, p>.15, \eta_G^2<.007$). 

Not surprisingly, the results suggest that when provided with objective feedback on trial-by-trial performance, a simplified model of reliability estimation (\textit{Accuracy} variant) was able to dissociate advisors based on their accuracy, but did not differentiate based on advisors' agreement profile. On the contrary, human data suggest that agreement influences trust even when trial-by-trial feedback is provided. More surprisingly, a model without access to objective accuracy but endowed with metacognitive insight (\textit{Confidence} variant) was also able to discriminate between equally agreeing but differently accurate partners. As shown in Table \ref{table:table_exp2}, the accurate agreeing advisor tends to agree more often than the inaccurate agreeing advisor when the participant is objectively correct (6.5 times out of 7 against 5.5 times out of 7) and less often when the participant is objectively wrong (1.5 thirds against 2.5 thirds). Trials when participants' initial judgment is correct are usually associated with greater confidence ratings \citep{Fleming2014a, Henmon1911, Koriat2012}, thus a strategy of reliability estimation relying on confidence can exploit this covariation to detect differences in accuracy, notwithstanding equal agreement rates. 

\subsection{Discussion}
Experiment 2 orthogonally manipulated the agreement rate and accuracy of advice presented to participants. Results show different behaviors according to whether or not participants (and the model) were provided with trial-by-trial feedback. Human participants, but not the model (\textit{Accuracy} variant), showed an effect of agreement even when given objective feedback on the advisors performance. When feedback was removed, participants' explicit trust reports showed an effect of agreement but not of accuracy, while their implicit confidence changes showed both effects. Similarly, a \textit{Confidence} model distinguished both dimensions of advice, namely agreement and accuracy, by scaling agreement with internal metacognitive signals. On the contrary, a \textit{Consensus} model showed a strong preference for agreeing advisors and showed only limited evidence of being able to distinguish accurate from inaccurate advisors. 

\paragraph{}
Unsurprisingly, advisor accuracy affected reported trust and influence measure in the Feedback group \citep{Behrens2008}. The finding that these participants also showed an agreement effect is, however, more surprising. The finding seems to suggest that agreement is perceived as a strong indicator of the reliability of advice, independent of the actual underlying accuracy. A possible explanation for this effect is that participants show a confirmation bias \citep{Nickerson1998} by which confirming evidence (in this case an agreement trial) is weighted more than disproving evidence (a disagreement trial). The \textit{Accuracy} model created to describe trust formation in cases when feedback is provided by the experimenter is thus not enough to describe participants' behavior as it fails to show a mediating effect of agreement. A combination of accuracy and internal confidence could perhaps explain the mechanism giving rise to participants' actual behavior.

\paragraph{}
When feedback was unavailable, both participants' implicit (influence) and explicit (trust) behavior is dominated by the agreement rate of the advisors rather than by their accuracy. One possibility is that the preprocessing performed on original questionnaire ratings through PCA to aggregate different questions together (see Supplementary Information) caused a loss of the information originally communicated by participants. This hypothesis was discarded after obtaining similar results when including only the first trust question (Q1), which directly asked participants to rate how "accurate" each advisor was. The dominance of agreement over accuracy might have resulted in a halo-dumping effect \citep{Clark1994} whereby, when prompted to discriminate among advisors, participants used only the most accessible dimension. Nevertheless, an effect of advisor accuracy was still observed in implicit confidence change behavior - and, as predicted by the \textit{Confidence} model - suggesting that some proximal cues to accuracy were available to them. It is unlikely that this cue was the agreement of the advice, as this dimension was orthogonal to accuracy by design. A more plausible hypothesis is that participants accumulated over time some internal uncertainty quantity (e.g. confidence) for each advisor separately. 

\paragraph{}
Collectively, these results demonstrate that agreement is a crucial cue to reliability, such that its effects are seen even when objective feedback is provided. The fact that agreement was such a strong predictor of advice impact on participants' opinions made it difficult to isolate the effect of internal confidence on advice evaluation. Experiment 3 was designed to prevent this issue and to match all advisors in terms of agreement and accuracy rates, while at the same time varying the amount of shared information and bias between the advisors and the participants. It provides a direct test of the hypothesis that confidence judgments are used to make useful inferences about others.

\section{Experiment 3}
Experiment 3 aimed to provide stronger evidence that subjective confidence contributes to the formation of judgments about advice reliability. In this experiment, we directly manipulated the probability that advice would agree with the participant's initial judgment, conditional on their initial confidence in that judgment. Three advisors were created by making advisors' judgments more or less biased towards the participant's confidence judgments in trials when the participant's initial judgment is correct: (1) an unbiased advisor who tended to agree with the participant's choice about 70\% of the time, independent of participant's confidence; (2) a bias-sharing advisor who tended to agree with the participant's initial choice particularly when the participant is confident; (3) an "anti-bias" advisor who tended to agree with the participant particularly when the participant is unsure. The labeling of the advisors in terms of bias does not reflect an actual bias manipulation, but rather reflects our aim to capture the patterns of agreement that would result from shared biases with the advisors. The manipulation is a special case of the general idea of judgment dependence among observers that was described above. This design is intended to capture a property that may hold in many real-world situations, where the opinions and choices of individuals are not independent, but instead are correlated (e.g., reflecting their shared reliance on common sources of information). Crucially, the effects of agreement-by-confidence were studied while matching the overall agreement rate and accuracy of the different advisors.

\subsection{Methods}
\subsubsection{Participants}
50 participants were tested and divided in the two experimental groups. Due to participants failing to attend or complete sessions, numbers across groups were unbalanced with 24 participants in the No-Feedback group and 26 in the Feedback group. Participants were recruited through local advertisement and the Oxford University online recruitment platform. Prior the start of the experiment all participants signed a consent form approved by the local ethical committee. Volunteers were compensated for their time via monetary compensation (10 \pounds /hour) or course credits.

\subsubsection{Paradigm}
The experiment consisted of 12 experimental blocks of 35 trials each, with each of the three advisors seen on 10 trials and with 5 null trials. The perceptual task was the same as for Experiments 1 and 2. Display parameters, input modality and stages within a trial remained unaltered. A different confidence rating scale was used in this experiment, because the 10-point scale used in Experiments 1 and 2 would not allow us to distinguish fine gradations in confidence that are needed in this design. Thus, in this experiment, participants rated their confidence on a 100-point scale (50 points per interval, left vs. right). All participants' stimuli were titrated to converge to 70.7\% decision accuracy. Two blocks of 25 trials served as the practice blocks and used a fourth practice advisor. Similar to Experiments 1 and 2, trust questionnaires were presented every two block and questionnaire data went through the same preprocessing pipeline.
  
\subsubsection{Manipulation}
\begin{figure}[H]
\centering
  \includegraphics[width=0.8\textwidth]%
    {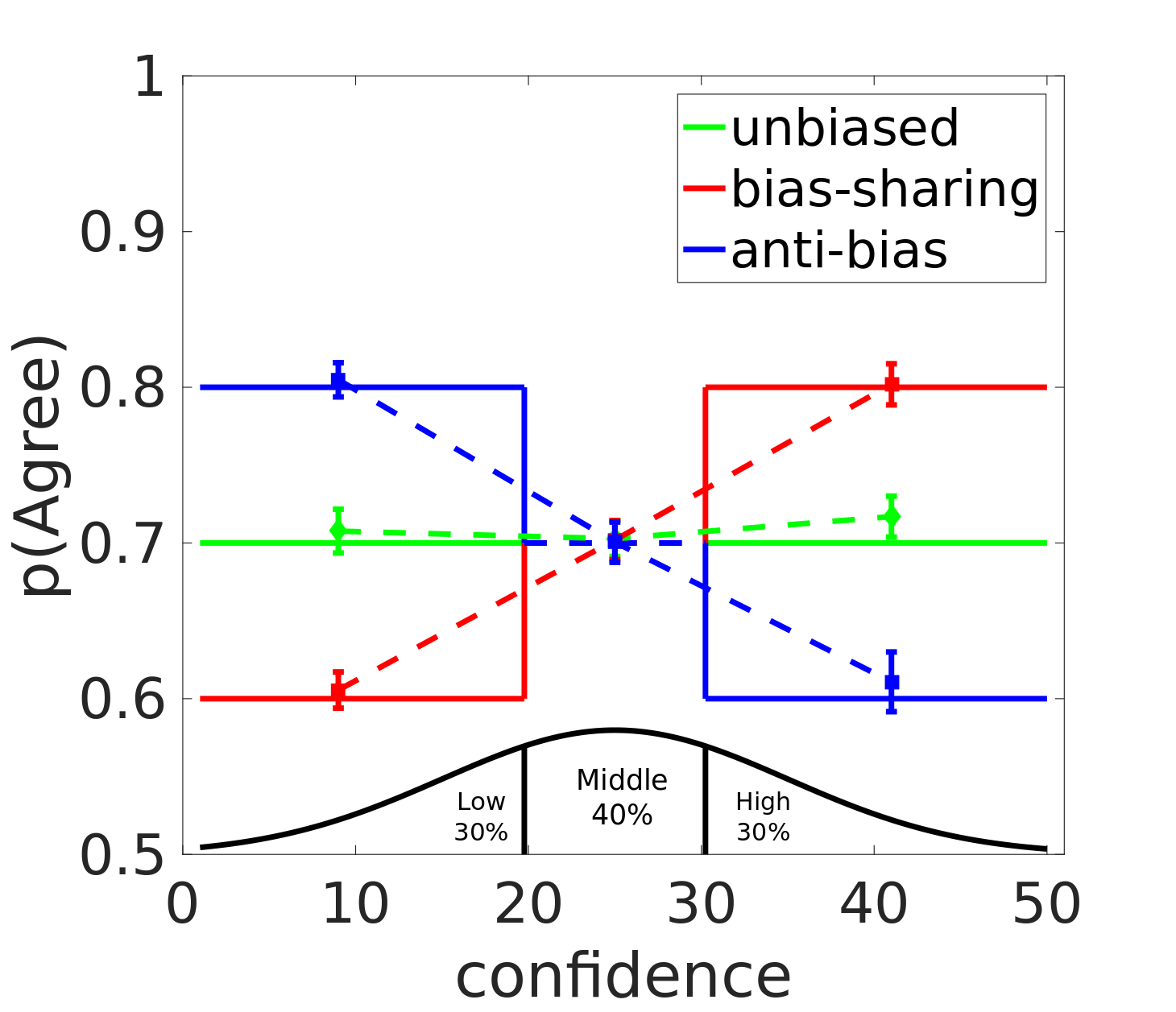}
  \caption[Experiment 3 manipulation.]{Experimental manipulation of Experiment 3. The probability of advisor's agreement conditional on participant's accuracy and confidence was manipulated so that the three advisors differed in their pattern of agreement (i.e., bias) despite being equal on average agreement rate and accuracy rate. Continuous lines represent expected agreement rates, dashed lines represent empirical data pooled across the two feedback groups.}
  \label{fig:exp3_paradigm}
\end{figure}

The advice profiles of the three advisors were manipulated so that all advisors showed on average the same level of accuracy and agreement rate with the participants' judgments. The pattern of agreement was manipulated, however, such that the three different advisors' likelihood of agreement with the participant varied according to the participant's initial confidence, as described in Table \ref{table:table_exp3} and illustrated in Figure \ref{fig:exp3_paradigm}. 

\paragraph{}
The distribution of the participant's pre-advice confidence judgments was divided into three confidence bins: the low, middle and high confidence bin, which include 30\%, 40\% and 30\% of the trials, respectively. On trials where the participant's pre-advice judgment was incorrect all advisors had a probability of agreement of 30\% independent of the participant's confidence level. On trials where the participant's judgment was correct, however, the three advisors had different agreement patterns. An unbiased advisor had a probability of agreement of 70\% independent of the participant's confidence. A bias-sharing advisor was defined as agreeing around 80\% of the time when the participant was highly confident and 60\% of the time when s/he was uncertain. An anti-bias advisor was designed to agree 60\% of the time when the participant was highly confident and 80\% of the time when s/he was uncertain. Importantly, however all advisors had an equal chance of agreement when the participant's pre-advice confidence fell in the middle bin (70\%). This ensured that all advisors were matched across all trials in terms of average agreement rate ($0.7*0.7+0.3*0.3 = 0.58$) and accuracy ($0.7*0.7+0.3*0.7 = 0.7$). 

\paragraph{}
By limiting analyses to trials within the intermediate confidence bin, we could compare advisors on trials that were matched for confidence and a priori likelihood of advice agreement. For example, asymmetric confidence changes in agreement and disagreement and larger advice influence in uncertain trials might contribute to advisor-specific distortions in the influence measure, if all trials were pooled together. The confidence reference distribution used to assign trials to the low, middle and high confidence bins was first set up on the basis of each participant’s confidence ratings in the first two practice blocks. The reference distribution was updated after each block to reflect the distribution of confidence judgments provided during the previous two blocks. This precaution was taken to take into account possible shifts of confidence during the course of the experiment.

\begin{table}[H]
\begin{tabular*}{\textwidth}{@{\extracolsep{\fill}}lrrr@{}} \toprule 
\multicolumn{1}{c}{} & \multicolumn{3}{c}{Advisors} \\ \cmidrule(r){2-4}
 & Bias-sharing & Unbiased & Anti-bias \\ \midrule
$p(Agree|Correct^s, Confidence_{low}^{s})$  & 60\% & 70\% & 80\% \\ \cmidrule(r){1-1}
$p(Agree|Correct^s, Confidence_{mid}^{s})$  & 70\% & 70\% & 70\% \\ \cmidrule(r){1-1}
$p(Agree|Correct^s, Confidence_{high}^{s})$  & 80\% & 70\% & 60\% \\ \cmidrule(r){1-1}
$p(Agree|Incorrect^s)$  & 30\% & 30\% & 30\%  \\ \midrule
$AUC(IG_e)$ & 14.63 & 14.74 & 15.78 \\ \bottomrule
\end{tabular*}
\caption[Experiment 3 advisors' profiles.]{Experiment 3 advisors' profiles. Agreement probability of different advisors is manipulated conditional on the participant's pre-advice confidence and accuracy. This manipulation allowed to create three different advisors who were matched in terms of agreement rate and accuracy, but who differed in terms of information as quantified by $AUC(IG_e)$, namely the area under the expected information gain curve (see Supplementary Information for details). }
\label{table:table_exp3}
\end{table}

Although average accuracy and agreement rates were matched across advisors, the information that participants could gain by receiving advice was not. In the present experiment, computing advisor's information gain from prior knowledge of their conditional probability was not possible as this was dependent on each participant's individual confidence distribution. However, according to a Bayesian normative model, greater changes along the confidence scale should take place when subjective initial confidence is low, and lower shifts should be observed when initial confidence is high. The anti-bias advisor is in this respect in the good position of giving supporting evidence (agreement) when it is needed the most (low subjective initial confidence). The bias-sharing advisor on the contrary tends to agree when their judgments are less effective (high subjective initial confidence). The intuition that the anti-bias advisor was the most informative was confirmed using a numerical simulation based on the nominal confidence updates of an ideal Bayesian observer performing the task (see Supplementary Information).

Following Experiment 2, we expected different patterns of results to emerge from feedback and feedback-free conditions. Specifically, based on the \textit{Confidence} model, we predict that when feedback is not available, trust reports and influence measure will favor the bias-sharing advisor over the other advisors. This follows from the fact that both participants and the \textit{Confidence} model will experience more high-confidence agreements with the bias-sharing advisor compared to the non-bias sharing one. This will happen even if advisors are matched in terms of accuracy and agreement, because participants and the \textit{Confidence} model will learn the association between internal metacognitive states (like confidence) with external contingencies (like agreements/disagreements). Neither the \textit{Consensus} or the \textit{Accuracy} models are expected to distinguish among advisors. When objective feedback is provided, we might expect that participants will not distinguish between advisors given that they are all equally reliable ($\sim$ 70\% accuracy).

The same measures of interest adopted in the first two experiment - namely the explicit trust measure and the implicit advice influence measure - were adopted here to compare how different advisors' profiles were perceived by the participants. Again, the three model variants were applied to the new data.

\subsubsection{Exclusion criteria}
An exclusion criterion based on staircase convergence was set so to exclude all participants who appeared to have a random guessing strategy. Application of this criterion resulted in the exclusion of one participant from the Feedback group and one participant from the No-Feedback group, leaving a total of 25 and 23 participants in these groups, respectively. Average difficulty parameter $d$ was $9.98 \pm 2.82$ (pooled data).

\subsection{Results}
The following analyses were performed to evaluate whether advisor types differed according to the measures of explicit trust and implicit opinion change. Furthermore we were interested in knowing whether this effect was modulated by the presence of feedback. Feedback acted as a between-participants factor while advisor type acted as a within-participants factor in the following mixed-design ANOVAs. For within-participants variables, degrees of freedom were corrected for violations of sphericity according to the Greenhouse-Geisser procedure, with epsilon values reported as appropriate.

\subsubsection{Trust ratings}
\begin{figure}[H]
\centering
  \includegraphics[width=1\textwidth]%
    {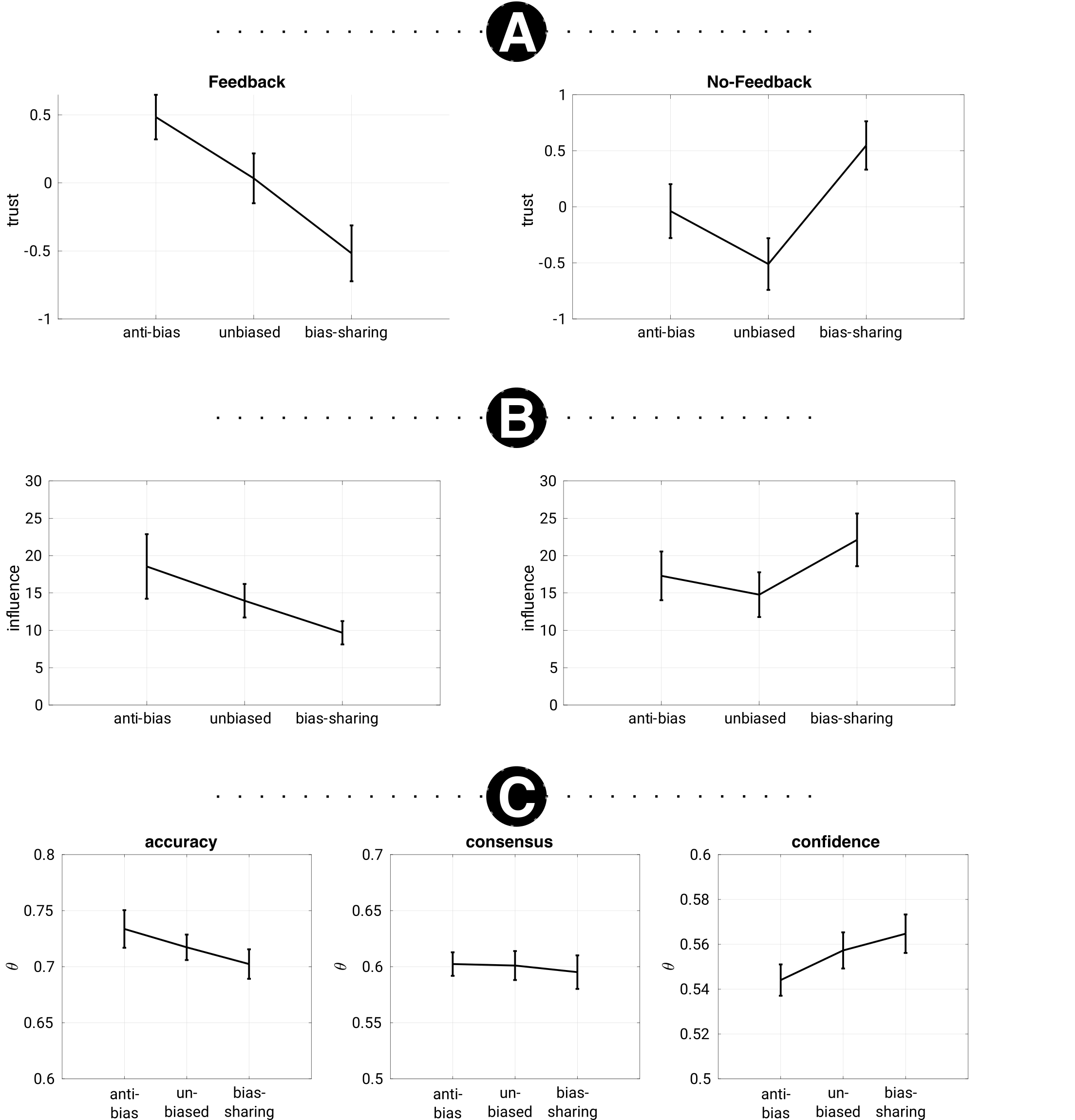}
  \caption[Experiment 3 - trust ratings and influence of human and simulations.]{Experiment 3 - trust ratings and influence of human participants and simulations. (a) The effect of advisor type on explicitly reported trust, separately for the Feedback and No-Feedback condition. Error bars represent s.e.m. (b) The effect of advisor type on the influence measure, divided by the two feedback conditions. Error bars represent s.e.m. (c) Average $\theta$-values, representing the model's trust in different advisors, differ depending on what pieces of information the model has access to. Error bars represent s.e.m.}
  \label{fig:exp3_all}
\end{figure}

A mixed-design ANOVA was run on explicitly rated trust scores from the end-of-block questionnaires, with Feedback as a between-participants factor and Advisor Type as within-participants factor. This analysis revealed no significant main effect of Advisor Type ($F(2,92)=1.66, p=.19, \eta_G^2=.03, \epsilon = 0.99$) or Feedback ($F<1$), but a significant interaction between these factors ($F(1,92)=6.64, p=.002, \eta_G^2=.12, \epsilon = 0.99$). The results suggest that although different advisors were similarly trusted on average, the effect was modulated by the presence of feedback. Figure \ref{fig:exp3_all}A shows how the effect of advisor was influenced by the presence of feedback. 

Separate within-groups analyses were carried out to understand differences among advisors present in each experimental group. For the group of participants receiving trial-by-trial feedback, a one-way repeated measures ANOVA (Figure \ref{fig:exp3_all}A) revealed a significant effect of Advisor Type ($F(2,48)=4.90, p=.01, \eta_G^2=.16$), with rated trust being highest for the anti-bias advisor, intermediate for the unbiased advisor, and lowest for the bias-sharing advisor. Follow-up pairwise comparisons indicated that the bias-sharing advisor was significantly less trusted than the anti-bias advisor ($t(24)=3.09, p=.004, d=1.07, B_{H[0,1]}=47.34$), with no reliable differences observed between the anti-bias and the unbiased advisors ($t(24)=1.60, p=.12, d=0.51, B_{H[0,1]}=1.67$) or between the unbiased and the bias-sharing advisors ($t(24)=1.56, p=.13, d=0.56, B_{H[0,1]}=1.83$). This pattern, with a high level of trust in the anti-bias advisor and lower in the bias-sharing advisor was not predicted (a priori, or by the models as described below), but is in accordance with the pattern of information gain of each advisor (Table \ref{table:table_exp3}), suggesting that participants' reported trust was sensitive to the information carried by each advisor, rather than their simple accuracy. Similar results are reported below for implicit trust, suggesting this pattern is robust to different trust elicitation methods.

For trust ratings from the No-Feedback group, a repeated measures one-way ANOVA also revealed a significant difference among Advisors ($F(2,44)=3.56, p=.03, \eta_G^2=.13$), but the pattern observed was very different. As predicted, rated trust was highest for the bias-sharing advisor when feedback was absent, numerically so compared with the anti-bias advisor ($t(22)=1.48, p=.07, d=0.53$, one-tail, $B_{H[0,1]}=1.74$) and significantly so compared with the unbiased advisor ($t(22)=2.81, p=.005, d=0.98$, one-tail, $B_{H[0,1]}=22.42$). The anti-bias and the unbiased advisors did not significantly differ from each other ($p>.1, d=0.41, B_{H[0,1]}=1.13$), but rated trust was, unexpectedly, numerically higher for the anti-bias observer than the unbiased advisor. 

\subsubsection{Influence}
Similar patterns of trust were apparent as measured implicitly via advisor influence (confidence change following advice). A mixed-design ANOVA on measured influence with Feedback as a between-participants factor and Advisor Type as a within-participants factor, revealed no significant main effect of Feedback ($F(1,46)=1.36, p=.24, \eta_G^2=.01$) nor Advisor ($F(2,92)=1.12, p=.33, \eta_G^2=.009, \epsilon=0.77$), but a significant interaction between the two ($F(2,92)=4.80, p=.01, \eta_G^2=.03, \epsilon=0.77$), indicating that the presence of feedback modulated the influence that different advisors had on participants' opinions (Figure \ref{fig:exp3_all}B).

When looking at influence in the Feedback condition only, a one-way ANOVA showed a marginally significant effect of Advisor Type ($F(2,48)=2.88, p=.06, \eta_G^2=.05$). Planned comparisons (two-tailed t-tests) showed that the bias-sharing advisor was less influential than both the anti-bias advisor ($t(24)=1.98, p=.05, d=0.54, B_{H[0,5]}=3.71$) and the unbiased advisor ($t(24)=2.26, p=.03, d=0.44, B_{H[0,5]}=6.56$). The anti-bias advisor was numerically more influential than the anti-bias advisor, but this difference was not reliable ($t(24)=1.10, p=.28, d=0.26, B_{H[0,5]}=1.47$). 

A one-way ANOVA on influence measured in the No-Feedback condition revealed a significant effect of Advisor ($F(2,44)=3.25, p=.04, \eta_G^2=.03$). Planned comparisons showed that the bias-sharing advisor was significantly more influential than the unbiased advisor ($t(22)=2.63, p=.007, d=0.46$, one-tail, $B_{H[0,5]}=13.62$) and numerically more influential than the anti-bias advisor ($t(22)=1.46, p=.07, d=0.29$, one-tail, $B_{H[0,5]}=2.06$). No significant difference was found between the unbiased and the anti-bias advisors ($p>.1, d=0.16, B_{H[0,5]}=1.07$), but the direction of the difference was the same as for explicitly rated trust, with the anti-bias advisor somewhat more influential on participants' decisions and confidence than the unbiased advisor. Thus, these results seem to suggest that the bias-sharing advisor was more influential than the other two when trial-by-trial feedback was not available.

\paragraph{}
In summary, as in Experiment 2, we find that participants' perception of advice reliability varies systematically according to whether trial-by-trial feedback is present or absent when advice is correlated with their own initial decisions. In particular, trust in advisors follows their relative informativeness when feedback is provided, but when feedback is absent participants tend to trust in and be influenced by advisors who share their judgment biases (i.e., in advisors who agree with their confidently held judgments).

\subsection{Model results}
The following simulations and analyses were performed to explore the differing patterns of trust across advisors that are expected from simple models of estimating advisor reliability. Differences in $\theta$-values observed among the three model variants show how same experiences of stimuli and advice can give rise to different trust judgments depending on the information an observer has access to. Simulations are also useful in the context of this experiment, because they allow us to check that our designs were controlled as intended (e.g., for agreement rates across advisors) even though we had less precise control over conditions because counterbalancing depended on an evolving estimate of participants’ confidence distributions. Figure \ref{fig:exp3_all}C shows the pattern of results that the three model variants produce. 

Both when the model has access to trial-by-trial feedback (\textit{Accuracy} variant), and when it only has access to past agreement (\textit{Consensus} variant), no significant effect of Advisor is observed ($F(2,94)<1.70, p>.18, \eta_G^2<.02$), nor is there a difference between the bias-sharing and the anti-bias advisor. These patterns are expected because the three advisors were matched for accuracy and agreement rates by design in this experiment. On the contrary, a \textit{Confidence} variant - which uses metacognitive information and past agreement (but lacked access to trial-level feedback) - showed a significant effect of Advisor ($F(2,94)=7.95, p<.001, \eta_G^2=.10$). Specifically, simulated trust was higher for the bias-sharing advisor than the anti-bias advisor ($t(47)=3.54, p=.001, d=.74, B_{[0,.05]}=114.64$), and higher for the unbiased advisor than the anti-bias advisor ($t(47)=2.99,p=.004,d=.57, B_{[0,.05]}=18.16$). Simulated trust was higher for the unbiased than the bias-sharing advisor, but this difference was not reliable ($t(47)=1.21,p=.22,d=.25, B_{[0,.05]}=0.41$). These findings indicate that by accessing metacognitive signals (as provided in the model by participants' confidence ratings) the model was able to discriminate among different advisors. This model correctly predicts greatest levels of trust in a bias-sharing advisor, but also predicts lowest levels of trust in anti-bias advisors, whereas our experimental participants expressed (numerically) lowest levels of trust in unbiased advisors.

\paragraph{Post-advice confidence correlations.} 
As described above, our main analyses for each experiment focused on qualitative predictions arising from different strategies for inferring advisor reliability. Collectively, the results are consistent with the hypothesis that people use their internal sense of confidence in making these inferences - showing sensitivity to advisor accuracy in the absence of objective feedback, even when advisors are matched for agreement rate, and developing differing patterns of trust when advisor agreement rates vary with their own expressed decision confidence. Our final analysis of the empirical data attempted a more quantitative comparison of model predictions, specifically focusing on the whether the \textit{Consensus} or \textit{Confidence} models better predicted post-advice confidence ratings across trials for the participants in the No Feedback conditions of Experiments 1-3.

For this analysis we used Bayes rule to infer the trial-by-trial post-advice confidence ratings that each variant would express given a participant's expressed pre-advice confidence and advisor agreement (see SI). The within-participant correlation between participants' post-advice confidence and model's post-advice confidence was computed for each experiment, for No-Feedback groups only. Second-order statistics were performed to test, across experiments, which variant was more strongly correlated with the human data. A 2x2 ANOVA on correlation coefficients with Model (\textit{Consensus} vs. \textit{Confidence}) and Experiment as factors showed that the \textit{Confidence} variant was significantly more correlated with participants' responses than the \textit{Consensus} variant ($F(1,22)=8.18, p=0.009, \eta_G^2=0.0049$). No significant effect of experiment nor interaction between the two were found ($F(2,44)<1.3, p>.25$), suggesting that, across experiments, the \textit{Confidence} model's post-advice confidence more strongly covaried with participants' true post-advice responses. As a check for the soundness of this model comparison method, the same 2x2 ANOVA was run on the correlation coefficients between participants' post-advice confidence and the model's post-advice confidence predictions, after randomly shuffling trials within each participant. This operation should ensure that any advantage of the \textit{Confidence} variant over the \textit{Consensus} variant is not due to unspecific factors (like being overall more conservative in updating confidence), but rather to trial-level variability. After reshuffling, the \textit{Consensus} and \textit{Confidence} variants were not significantly different from each other ($F(1,22=1.95, p=0.17, \eta_G^2=9.54e-04$), corroborating our conclusions.

\subsection{Experiment discussion}
Experiment 3 showed again that systematic differences in trust emerge as a function of feedback when advisors' judgements are non-independent from the advisee, and further demonstrated a strong influence of our agreement-in-confidence manipulation on trust. Here, the presence of feedback partly reversed the pattern of trust ratings and influence measure that was observed when trial-by-trial feedback was unavailable: When objective feedback was provided, people trusted more and were more influenced by an advisor who more frequently agreed with them when they themselves were unsure (vs. less frequently when they were sure), and showed less trust in an advisor who tended to agree with them in decisions in which they were already sure. On the contrary, when objective trial-by-trial feedback was removed, this pattern partly reversed such that trust and influence were greatest for the advisor who tended to agree with participants' confidently made judgments. Participants expressed less trust in an advisor who tended to disagree with their confidently made judgments and even more so in an unbiased advisor whose tendency to agree or disagree did not vary with the participants’ initial decision confidence.

\paragraph{}
Participants in the Feedback group showed an effect of advisor type even though the feedback indicated that all advisors were equal in terms of both agreement rate and accuracy. These findings contrast somewhat with the results of Experiment 2, where ratings of trust and measures of advice influence were dominated by agreement rates when these varied markedly across advisors. Evidently, though agreement appears to be a critical determinant of trust (with or without feedback), participants remain sensitive to other dimensions of advice. Here, they appeared sensitive to the informational value ("information gain", see SI) of the advice and not only to its reliability \textit{per se}: They trusted and were more influenced by advisors whose judgments were less redundant with their own, in particular showing greatest trust in the advisor whose advice was most likely to improve their performance (by reinforcing, through agreement, initial judgments that were correct but most likely to waver due to low initial confidence). 

As predicted, participants in the No Feedback group showed greatest trust in advisors who tended to agree with their own confidently expressed judgments. However, more surprisingly, participants did not show least trust in the anti-bias advisor who agreed with them more frequently when their initial judgment was made with low confidence. If anything, trust was lower in the unbiased advisor. The difference was not reliable and hence must be interpreted with caution, but we note a possible link to an existing proposal about how expertise is assessed in the absence of feedback. Thus, according to \cite{Weiss2003}, scarcely differentiated judgments across different observations tend to be indicative of lower expertise. In the present experiment, the unbiased advisor was characterized by this lack of differentiation, in that her advice agreement rate was constant as a function of participants' initial decision confidence, in contrast to the two other advisors. Nonetheless, regardless of the explanation of this unexpected result, the findings of Experiment 3 demonstrate participants' sensitivity to advice in relation to their own confidence in their judgments, and learn differentiated patterns of trust accordingly.

\section{Agent-based modeling: network-level impacts of individuals' trust strategies}
The experimental and simulation results described above demonstrate that the absence of trial-level objective feedback does not preclude agents from inferring the trustworthiness of other agents. Simple heuristics based on agreement and decision confidence can be good proxies for objective feedback. However, our findings also highlight that systematic deviations between feedback and feedback-free scenarios - and, correspondingly, systematic biases in trust formation - can be observed when judges' and advisors' judgments are correlated (e.g., due to shared biases or common sources of information). Thus far we have focused on the situation of an individual decision maker learning about the quality of advice from different sources. To extend this work, we explored how these effects might play out in more complex, multi-actor situations using agent-based modeling. 

Using an agreement (or agreement-in-confidence) heuristic can be an effective strategy to infer partners' reliability when more objective cues are not available. However, predictable distortions are expected when independence is not preserved among judges. If patterns of trust vary depending on feedback availability, we might expect different macro-level patterns of trust in networks where feedback is available rather than difficult to obtain. Furthermore, the presence of bi-directional information channels between agents (as opposed to judge-advisor  systems) may lead to clustering of people who share the same biases, increasing the polarization of their views (i.e., confidence magnitude) as well as producing phenomena of \textit{groupthink} \citep{Janis1972, Turner1998}. These effects parallel important observations about human social networks. Informational echo-chambers - whereby individuals with similar characteristics tend to form clusters that are impenetrable to external information - have been often described in social sciences \citep{Sunstein2001,Jasny2015}. When exposed to competing opinions on social media, people belonging to the same cluster may prefer, consume and share within-cluster information more than between-cluster information \citep{Bessi2016,DelVicario2016}, in a similar manner that people in a group prefer to discuss shared information than private information \citep{Schulz-Hardt2006,Stasser2003}. If so, the present ideas might provide a normative account of the origin of these effects that otherwise seem suboptimal: According to our findings, preference for agreeing opinions (particularly those that agree with our confidently held views) is an adaptive strategy for estimating advice reliability in the absence of an objective standard, but can lead to these systematic biases in trust when opinions and judgments are non-independent.

To explore this hypothesis, we used agent-based modeling to explore how the manipulation of feedback availability affects network structure and information flow. Agent-based models are useful tools to study phenomena of emergence from simple interacting parts. Manipulating agents' access to metacognitive signals can thus shed light on how the cognitive structure of agents can affect the structure of a network \citep{Epstein2013}. Giving agents choice over what information to sample - for example, if reliability estimates were used to guide advisor choice \citep{Denrell2005,Aiello2012} - might further remove agents from being exposed to useful but uncorrelated information (cf. our "anti-bias" advisor in Experiment 3). 

In the following section, we test how macro-scale patterns of trust and beliefs are affected by the presence or absence of feedback in networks of agents employing the cognitive strategies described so far in this work. We expect that (a) When judgments in the population are independent, agreement-based heuristics are useful in reliably approximating true underlying expertise; (b) when judgments are correlated - e.g., due to the presence of shared biases in the population - agents sharing similar biases will tend to segregate due to the use of a confidence-agreement heuristic in feedback-free (but not feedback-rich) scenarios; (c) The use of trust to discount advice will be beneficial to agents' accuracy when judgments are independent but not when they are correlated. 

\subsection{Model description}
An agent-based model was programmed using NetLogo \citep{Wilensky1999} and is available upon request at \url{ https://github.com/chri4354/trust\_formation\_without\_feedback}.  It simulates development of trust among agents as they make decisions, and revise these decisions on the basis of learning about the opinions of others, in a series of simple binary choices: Is a stimulus from category A or B, as a generalised case of the specific dot discrimination task used in our experiments. Of interest was how trust among agents was affected by the relative quality of their decisions, the availability of feedback and, crucially, the degree to which agents shared biases in their decision processes. Our models formalised these biases as differences in prior (base rate) expectations about the likelihood of A or B being the correct answer, which via Bayesian belief updating will bias the interpretation of incoming information in the decision process (as an analogue of what might be expected to happen in everyday decisions as a function of our political leanings, musical preferences, tastes in wine, etc.).

The model was initialised as a fully connected directed network of $N$ agents. A directed edge from agent $i$ to agent $j$ represents the trust $\theta_{i,j}$ that $i$ has in $j$'s opinions. We simulate agents on a lattice network performing repeated binary A/B decisions, receiving advice from other agents, inferring their reliability and updating their own initial decisions. We let the simulation run for a 1000 steps. A signal $s$ with strength $S$ was drawn from a uniform distribution between $- \frac{S}{2}$ and $+\frac{S}{2}$. This represents the decision quantity to estimate (e.g., difference in dots or true state of the world). The task of each agent was to determine if $s$ was positive (event A) or negative (event B). Each agent estimated the posterior probability of $A$ given the perceptual information generated by $s$ as follows:

\begin{equation}
    p'(A) = p(A|E_p) = \frac{p(A)E_p}{p(A)E_p + p(\bar{A})\bar{E_p}}
    \label{perceptual_posterior}
\end{equation}

\begin{equation}
    E_p = L(s + \mathcal{N}(0,\sigma))
\end{equation}

where $p(A)$ is the prior probability of observing $A$s before seeing any stimulus, $L$ is a logistic sigmoid mapping from sensory evidence to probability; $E_p$ is the perceptual evidence resulting from such mapping and $\mathcal{N}$ is independent individual perceptual Gaussian noise with mean 0 and standard deviation $\sigma$. Each agent's perceptual noise (and thus accuracy) was manipulated by varying the noise parameter $\sigma$. Each agent's bias was manipulated by varying the value $p(A)$. Agents' confidence was represented as the distance from the uncertainty point $.5$:

\begin{equation}
    C = .50 + \lvert {p'(A) - .50}\rvert
\end{equation}

Trust, represented by the network's edges, was initialized to 0.50 for every agent and updated after social interaction. After making a judgment, agents selected one other agent to interact with either at random (i.e., without biased sampling) or proportionally to their trust (with biased sampling). Agents then updated their initial judgment $p'(A)$ as follows:

\begin{equation}
    \hat{p}(A) = p(A|E_s) = \frac{p'(A)E_s}{p'(A)E_s + p'(\bar{A})\bar{E_s}}
\end{equation}

where $E_s$ represents social evidence and is obtained from the advisor's judgment either by taking the advisor's raw judgment $p'(A)$ (without advice discounting) or by discounting the advisor's judgment proportionally to the agent's trust in the advisor (with advice discounting). Advice discounting consisted in a linear regression toward the uncertainty point .5. (equation S20). After updating their judgments, each agent $i$ updated its trust judgments (i.e., outward edges) based on the available information about other agents. If feedback is available, the agent updates its current trust in agent $j$ by virtue of a delta rule in the form:
\begin{equation}
    \theta_{i,j}^{t+1} = \theta_{i,j}^t + \alpha (F_j - \theta_{i,j}^t)
    \label{trustUpdate}
\end{equation}

where $F_j$ is the accuracy of agent $j$ and $\alpha$ is a learning rate set to 0.1. If feedback is not available on the contrary, the agent replaces $F$ with $\hat{F}$, or the \textit{estimated} partner's accuracy. $\hat{F}$ was calculated using the agreement or agreement in confidence heuristics described above. In our simulations, we assessed the effect of feedback availability as it varied parametrically, from being available after every decision (i.e., p-feedback = 1.0), available after only some decisions (i.e., 0 < p-feedback < 1), or never available (i.e., p-feedback = 0), rather than the simpler case of feedback presence/absence that we studied experimentally above.

The emergence of trust patterns when using agreement-based heuristics can be expected to track true accuracy when judgments are independent but generate clustering of populations when judgment correlations emerge within such populations. We defined a network's clustering coefficient as the ratio between average trust toward agents who initially share the same bias (in-group trust) and total average trust: $\Bar{\theta}_{in-group} / (\Bar{\theta}_{in-group} + \Bar{\theta}_{out-group})$. A ratio of 0.5 represents no preference (i.e., no difference in trust) toward agents sharing the same bias, while a ratio greater than 0.5 represents a preference toward agents sharing the same initial biases. We test how network clustering is shaped by the presence or absence of objective feedback and show that bias-specific segregation arises only in the absence of feedback. 

Finally, once bias-specific segregation is established, we ask whether such clustering remains stable. In particular, after 500 iterations we allow agents to dynamically change their original bias as a function of experience. For example, it is known that the bias observed in people performing binary judgments is influenced by their recent history of decisions \citep{Akaishi2014,Zylberberg2018}. In the present context, if an agent systematically reports "A" but receives negative feedback, they should reduce their bias by decreasing their prior probability $p(a)$. Similarly, when feedback is absent, an agent who systematically reports "A" but finds themselves, after interacting with other agents, believing that $B$s are more frequent than expected, should reduce their bias towards $A$s. Conversely, bias should get stronger if the social contexts reinforces it (although see \citep{Bail2018}). We modelled bias update with a delta rule: 

\begin{equation}
    p(A)^{t+1} = p(A)^t + \alpha(I^t - p(A)^t)
\end{equation}

where $I^t$ is an indicator variable that represents the final belief in the event $A$. When objective feedback is available, $I$ takes the value of 1 if an event $A$ occurred and 0 otherwise. When feedback is not available, $I$ is set to the discrete or continuous final subjective belief in the event $A$. In the following section, we show the results obtained when a discrete final belief is used in the absence of feedback: 
\[
    I = 
\begin{cases}
    1,& \text{if } \hat{p}(A)\geq 0.5\\
    0,              & \text{otherwise}
\end{cases}
\]

Similar results were obtained setting $I$ to the continuous belief $\hat{p}(A)$.

\subsection{Results}
\begin{figure}[H]
\centering
  \includegraphics[width=0.8\textwidth]%
    {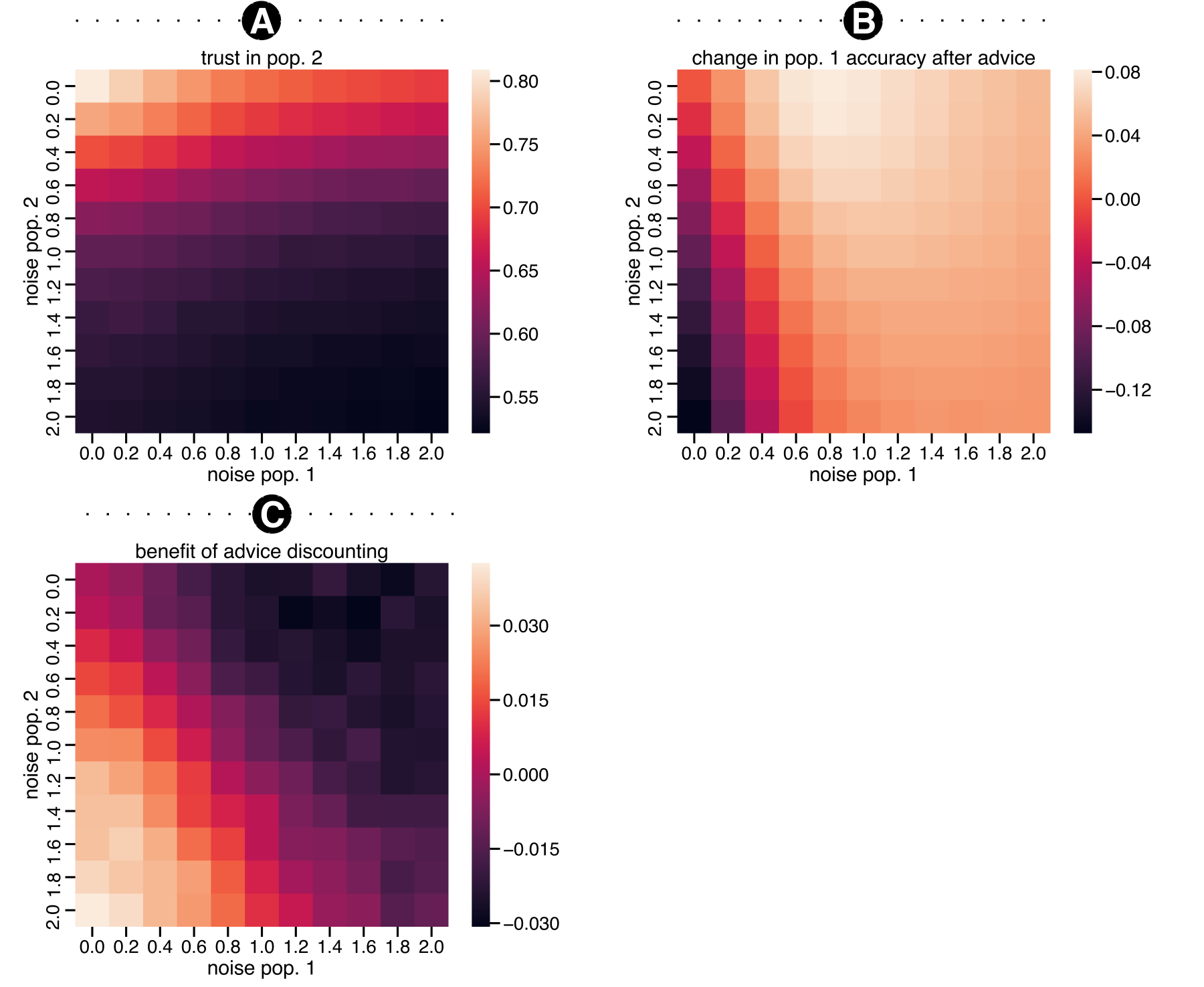}
  \caption[Agent-based model - trust]{A: average trust shown in agents belonging to population 2. Trust in these agents increases as the noise affecting them decreases. Interestingly, trust formation is affected by perceptual noise of the observed agent(s) as well as perceptual noise of the observing agent(s) [cf.\citep{Kruger1999}]. B: benefit of advice for Population 1. Pixels' color represent the difference between post and pre-advice accuracy. C: the use of a trust-based advice discounting strategy increases post-interaction accuracy of accurate agents' but not inaccurate ones.}
  \label{fig:ABMtrust}
\end{figure}

We first tested the hypothesis that agreement-based heuristics are adaptive in situations of independent judgments because they reliably track others' expertise without the need to rely on external forms of feedback. To this end, we set all agents' initial bias $p(A)$ to 0.50 and the probability of feedback to 0. We then created two sub-populations of agents that orthogonally varied in their overall judgment accuracy (modeled as the degree of perceptual noise, with decreasing accuracy as a function of increasing noise). We then calculated the average trust toward a target sub-population (here, Population 2). The results show that average trust in Population 2 agents correctly tracks their underlying perceptual noise: Trust in Population 2 is highest when their judgments are more accurate (i.e., when their perceptual noise is low: upper pixels in Figure 7A) and decreases as their perceptual noise increases. Interestingly, the effect is not linear and we observe an interaction between the accuracy of Population 1 and the accuracy of Population 2 (Figures \ref{fig:ABMtrust} upper panel, S3), mirroring the simple numerical analysis shown in Figure \ref{fig:paradigm}B: In the absence of feedback, poor performers in Population 1 fail to distinguish the accuracy of others, evident as a narrower range of trust values in rightward pixels in Figure 7A [cf. \citep{Kruger1999}]. Nevertheless, overall, we see that trust in Population 2 systematically tracks its level of performance, even when agents lack any objective standard against which to judge their own decisions or those of others.

Our second hypothesis was that when the true state of the world is difficult to discern (e.g., when objective feedback is absent, signal strength is weak or perceptual noise is large), bias-specific segregation can arise. As described above, we quantify this segregation in terms of a \textit{clustering} metric, which indicates the degree to which trust in in-group members exceeds trust in out-group members (indicated by clustering $>$ 0.5).

\begin{figure}[H]
\centering
  \includegraphics[width=0.7\textwidth]%
    {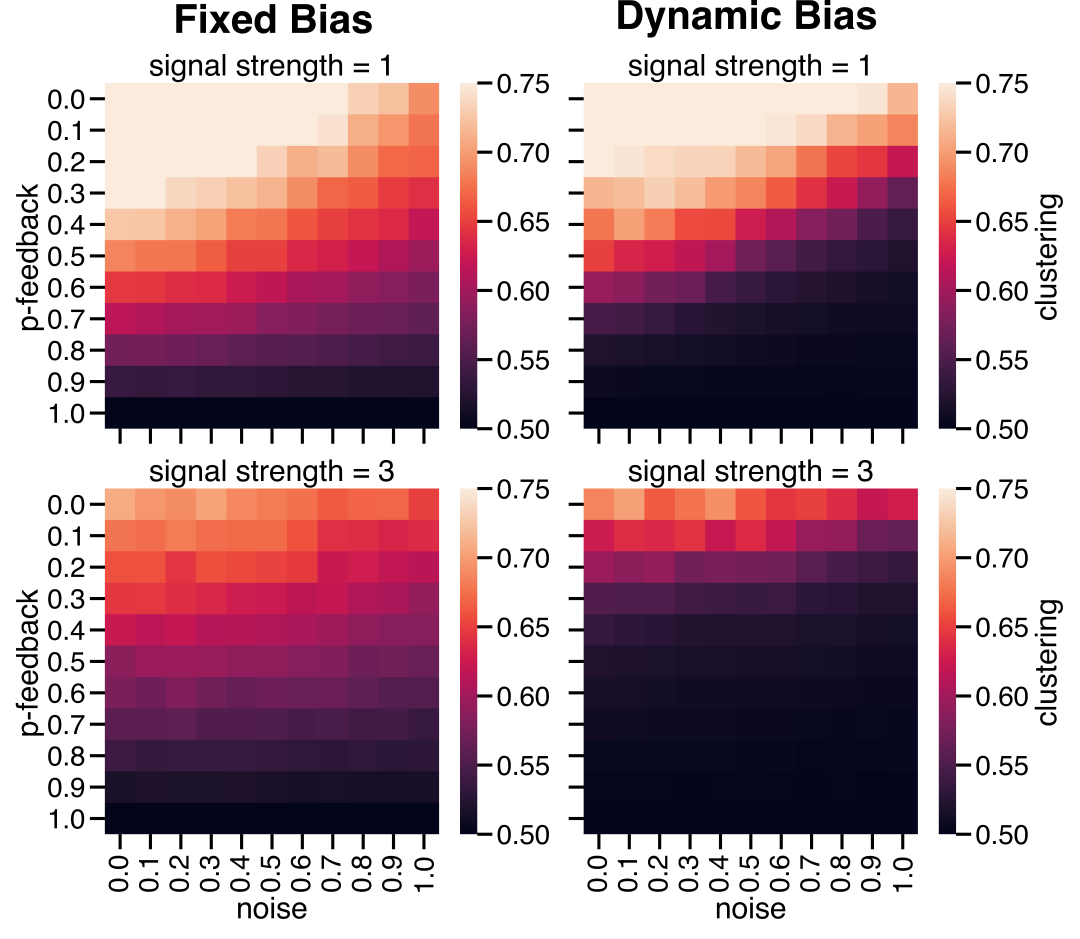}
  \caption[Agent-based model - Results]{Clustering as a function of signal strength, probability of feedback and noise. Each pixel in each panel indicates the average clustering value (over 20 simulations), computed as described in the main text, after 1000 iterations of decision, advice and update in a network of agents with fixed stimulus strength, perceptual noise and feedback probability parameters. Right and left columns distinguish simulations with and without bias update respectively. Rows from top to bottom represent increasing signal strengths.}
  \label{fig:ABMclustering}
\end{figure}

In this simulation, we varied the bias $p(A)$ shown by two subpopulations of equal size and equal average accuracy (equivalent average perceptual noise). For each individual, we randomly draw $p(A)$ from a uniform distribution in the range $[0,0.50]$ (Population 1) and $[0.50,1]$ (Population 2): That is, the two populations differ in their base rate estimates of the relative likelihood of the two outcomes (A and B), with Population 1 biased towards judging events as "A"s and Population 2 biased towards judging events as "B"s. On each iteration, agents interacted with another randomly selected agent and updated their trust according to equation \ref{trustUpdate}. Figure \ref{fig:ABMclustering} shows the average clustering shown after 1000 iterations as a function of signal strength (from low in the upper panels to high in the lower panels), perceptual noise (variation along the x-axis of each panel), feedback probability (y-axis), and whether agents update their bias depending on recent experience (left column: without bias update; vs. right column: with bias update).

Average clustering decreased as the probability of receiving objective feedback increased, evident in Figure \ref{fig:ABMclustering} as darker pixels in the lower parts of each panel. When objective feedback is consistently presented, trust is determined by an advisor's objective accuracy rather than their agreement with the agents' own judgments on each trial. Trust therefore converges to the same value for all agents, regardless of group, reflecting the fact that underlying accuracy (dependent on the level of perceptual noise) was equal on average across the entire population. A similar principle explains why clustering also decreased as signal strength increased - evident in Figure \ref{fig:ABMclustering} as darker pixels on average in the lower panels of the figure than in the upper panels: When signal strength is high, agents' decisions are dominated by external evidence rather than their prior expectations (as per Equation \ref{perceptual_posterior} above), thus agreement levels are high with both in-group and out-group members, and agents learn similar levels of trust in others across the whole population. On the other hand, clustering tends to decrease as perceptual noise increases (i.e., as agents' average accuracy decreases), evident in Figure \ref{fig:ABMclustering} as darker pixels to the right-hand side of each panel. This observation reflects the fact that as agents' accuracy tends towards chance levels, their likelihood of agreement with others will correspondingly tend towards chance (regardless of others' accuracy) (Figure S6), and they therefore lose the ability to distinguish advisor accuracy in the absence of an objective standard (cf. Figure \ref{fig:paradigm}B), and learn low levels of trust in both in-group and out-group members (Figure S5). 

On the contrary, agents in each sub-population tended to trust in-group members more than outgroup members as a function of decreasing feedback availability and signal strength, with this dependence disrupted under conditions of increasing perceptual noise. When signal strength is low, agents’ decisions are more strongly influenced by their prior expectations (as they should be, according to Bayes theorem). As a consequence, when feedback is infrequent or absent, agents’ trust is dominated by agreement over objective accuracy, and clustering with like-minded (bias-sharing) in-group members results, a function of increased agreement rates among members of homogeneous populations. This finding suggests that a simple delta rule of trust update can perform very differently depending on what source of information is used to supervise learning. The circular nature of using one's own belief to estimate others' accuracy produces higher clustering and segregation in scenarios where feedback is infrequent or absent, and decisions are often ambiguous. Meanwhile, our results are also in accordance with recent studies in ecology and network science, which suggest that the presence of noise is beneficial to group decision making to the extent that it reduces the risk of getting stuck on local optima \citep{Couzin2011, Kao2014, Shirado2017}: Here, high levels of noise disrupt the formation of clustering according to pre-existing biases (albeit by reducing trust in both in-group and out-group members alike) (Figure S5).

Our third and final hypothesis was that segregation between in-group and out-group members, once established, can resist modification and in turn affect the persistence of shared beliefs. In this simulation, agents first made decisions with fixed biases (i.e., a fixed p(A) value) and learnt trust in other agents across multiple iterations, as in the previous simulation. As before, the two populations drew their initial biases from random distributions favoring either A or B as a prior (p(A) $>$ .5 or p(A) $<$ .5, respectively), with feedback availability and perceptual noise level fixed in each simulation run but varied across runs to explore the impact of these parameters on network behavior. Signal strength was fixed at 1, where a wider range of behaviors was previously observed (Figure \ref{fig:ABMclustering}). However, crucially in this simulation, after 500 iterations we allowed agents to update their bias, whereby their experience of the relative prevalence of A and B outcomes (as determined by their belief-influenced interpretation of sensory evidence, opinions of partners, and objective feedback where available) leads to modification of their estimates of p(A). Of interest was the way in which the biases of the two groups evolved across interactions as a function of feedback availability, perceptual noise, and information sampling strategy. 

\begin{figure}[H]
\centering
  \includegraphics[width=\textwidth]%
    {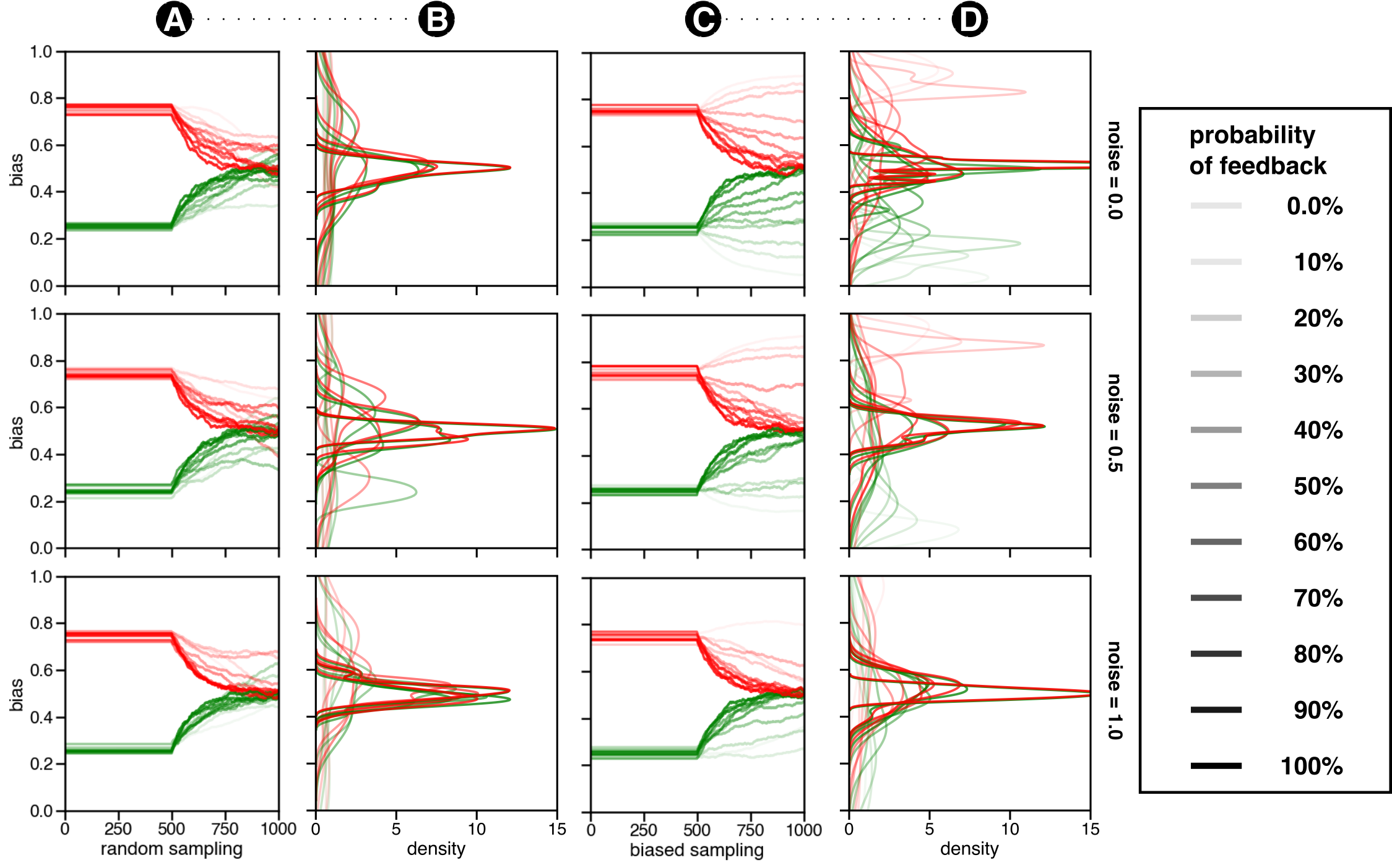}
  \caption[Agent-based model - Bias]{Bias as a function of time, feedback probability and agents' perceptual noise. Rows represent increasing levels of noise. Lines alpha values represent probability of objective feedback. Columns A and B show agents' bias evolution over time and final bias distribution when agents choose their partners at random. Columns C and D show the same graphs when agents choose their partner proportionally to their trust. Color saturation represents feedback probability and color represents the two original bias subpopulations.}
  \label{fig:ABMbias}
\end{figure}

The results are shown in Figure \ref{fig:ABMbias}, which plots the results of separate simulations for populations of agents that select partners either at random (columns A and B) or according to their current levels of trust in other agents (columns C and D). Columns A and C plot bias (i.e., mean p(A) value) across 1000 iterations of the decision task for the two groups in the population - one initially biased toward A decisions (red lines) and the other initially biased toward B decisions (green lines). Columns B and D plot the distribution of biases across members of each group after the final (1000th) iteration.

As shown in Figure \ref{fig:ABMbias}, when feedback is always available (saturated red and green lines), decision biases rapidly diminish such that agents from both populations converge on accurate estimates of the relative likelihood of A vs. B outcomes (i.e., unbiased estimates that vary narrowly around p(A) = 0.5). This convergence occurs irrespective of the level of perceptual noise and the way agents select their partners, reflecting the obvious value of having an objective standard against which to evaluate decisions. However, as feedback becomes less and less available (indicated in Figure \ref{fig:ABMbias} as reducing saturation of line colors), population biases are more persistent in a manner that exhibits sensitivity to both information selection strategy and levels of perceptual noise. Consider first the simulations of populations with random sampling of partners (columns A and B). When agents randomly sample their advisors, they tend to converge on accurate estimates of p(A) (i.e., $\sim=0.5$). This occurs even when feedback is not available (as indicated by low saturation lines) and independently of noise level (as indicated by different rows). The reason is that when agents sample their advisors at random, agents are equally likely to receive advice that the correct answer is A vs. B (or in other words advice that is characterized by their same bias vs. opposite bias). However, when agents weight advice in proportion to trust, they are more strongly influenced by other agents who share their initial bias (and therefore in whom they learn more trust based on patterns of agreement, as explored in the previous simulation) (Figure S4). Thus, social information can stabilise initial group biases insofar that agents rely on this social information as feedback probability decreases (populations with lighter colored lines exhibit systematically larger biases in Figure \ref{fig:ABMbias} and S4). The lower panels of Figure S4 A-B show that these population biases are disrupted when agents have higher levels of perceptual noise: This effect is partly due to noise reducing the influence of prior expectations on individual agents own decisions, but is also due to noise disrupting trust in other agents who share an initial bias (because noise will tend to reduce agreement rates with these individuals (Figures S5 and S6)).

As shown in the right hand panels of Figure \ref{fig:ABMbias}, the stabilising effect of social information also shows up when agents select partners to receive advice from as a function of their trust in those agents (columns C and D). Under these circumstances, stable group biases can persist even when feedback probability is as high as 0.8 (in the simulation when noise is absent, Figure \ref{fig:ABMbias}C, top panel), even if those biases are relatively small. This stabilisation is a natural consequence of preferentially seeking advice from agents who share a bias toward believing p(A) is systematically greater than (red group) or lower than (green group) its true value. Perhaps even more strikingly, the simulations also reveal cases where social information leads to extreme biases when objective feedback is rarely available: In these cases (light colored lines, e.g., p(feedback) $<$ .4 in the top panel of Figure \ref{fig:ABMbias}C), group biases actually become more extreme over time. This phenomenon emerges from a positive feedback loop whereby agents who share an initial bias (of, say, believing that p(A) $>$ 0.5) will tend to ask each other for advice more often, thereby compounding each others biased beliefs, and thereby increasing their tendency to seek advice from like-minded agents, thereby compounding their bias, and so on across iterations. The two groups therefore diverge to separate stable networks that exhibit extreme biases (of p(A) $\sim=$ 0 and p(A) $\sim=$ 1) and a strong tendency only to trust and seek advice from in-group members. In this way, sub-optimal behaviors at the network-level - such as self-sustaining biases and even extremes of polarization - can emerge from decision and learning strategies that are normatively sensible at the level of individual agents: Using decision confidence as a proxy for objective accuracy can allow agents to learn about the reliability of social information sources when feedback is infrequent or even absent. Moreover, using social information can in turn allow agents to learn about the world under these conditions of diminished feedback. However, these learning strategies exhibit important limitations when agents’ initial decisions are not independent but rather exhibit patterns of shared vs. distinct biases across individuals within a population.

Overall, the results of the agent-based modelling show the macro-scale effects of feedback availability, use of different reliability estimation strategies (feedback vs. agreement-based), and partner-selection strategies. Contrary to judge-advisor systems - where information flows from the advisor to the judge - here we allow agents to share information bidirectionally and use this information to estimate each other's reliability. Once again, agreement-based strategies provide an effective way to learn the reliability of social partners when feedback is not available and judgments are independent. However, they lead to predictable patterns of clustering and trust when agents' agreement rates are inflated by their shared biases or access to correlated information. Moreover, bias updating and partner-selection strategies also affect the structure of the network by changing the segregation of individuals sharing the same bias, leading to stabilisation of networks with systematically biased or even extremely polarized beliefs about the world (in our simulations, the relative likelihood of A vs. B being the correct decision).

\section{General Discussion}
The present research explored how people learn the reliability of social information sources in contexts where objective external feedback is not readily available. In these scenarios, we hypothesize that metacognitive confidence provides a useful proxy for objective feedback. This is due to the fact that in many tasks, confidence provides a finely calibrated estimate of underlying accuracy \citep{Pescetelli2016, Koriat2012a, Henmon1911, Fleming2016} as it represents the subjectively estimated likelihood of a correct decision being made \citep{Fleming2017, Aitchison2015}. Thus, people can apply this internal confidence estimate to received advice in order to estimate its reliability (as p(correct) in the case of agreement, and 1-p(correct) in the case of disagreement). Experiment 1 empirical data and simulations show the power of this agreement-in-confidence heuristic in identifying subtle differences in advice quality when advice provides new independent information: Participants (and simple models) learned to distinguish advisors of differing accuracy and confidence calibration even in the absence of objective feedback, and did so to a similar extent to participants (and simple models) who had access to objective feedback after every decision. Thus, we propose that people are able to follow internal signals (e.g., decision confidence) in a comparable manner to external signals - such as rewards or feedback - to learn about the reliability of advice and advisors according to associative learning principles \citep{Behrens2008,Sutton1998}. 

These findings extend our understanding of the uses of metacognitive signals. The simulation results show that a model endowed with access to its own metacognitive signals can discriminate between advisors better than models that only have access to objective feedback or agreement rate. Thus, metacognitive signals like confidence judgments are useful not only for internal uncertainty monitoring \citep{Yeung2012}, information seeking \citep{Desender2018}, cognitive control \citep{Botvinick2001} and social coordination \citep{Bahrami2010, Shea2014}, but are also useful in evaluating the reliability of external information sources. The present findings demonstrate the potential bi-directionality of this inference process: Confidence is not only the end-product of information flow from the external stimulus to a perceptual inference, but it can feed back to help making inferences about external events. Confidence represents a probabilistic estimation that once formed can be used to infer the state of variables that did not directly generate it. Evidence seems to support this notion, as confidence signals have been shown to parallel reward prediction error signals in feedback-free situations \citep{Zylberberg2018, Guggenmos2016}.

Our findings also extend previous work on the role of confidence in group decision making \citep{Sorkin2001,Bahrami2010, Koriat2012}. Previous studies have documented the way that the confidence accompanying advice provides a useful and impactful signal of its reliability \citep{Bahrami2010, Swol2005}. Here, we explore the converse situation of how an advisee's internal sense of confidence can modulate perceptions of advice. Thus, whereas previous studies have shown how confidence judgments from two individuals can be used to combine optimally the individual estimates of external stimulus attributes, our experiments and simulations wants to show that an observer's internal confidence about the state of the external stimulus can also be used to estimate the reliability of a social partner over and beyond what can be inferred from objective external cues. We used a traditional Judge-Advisor System paradigm \citep{Bonaccio2006,Sniezek1989,Yaniv2000} but incorporated a perceptual decision making paradigm to allow precise control over important variables such as strength of evidence available for the initial decision, the information carried by advice, and the manipulation of objective feedback to enable social decision making to be understood in terms of current theories of confidence as a probabilistic computation \citep{Pouget2016}.

\paragraph{}
Experiments 2 and 3 explored crucial limitations of the agreement-in-confidence heuristic, which leads to systematic biases in trust and influence when advice is not independent of an individual's own opinions. Much of the work in psychology in the last decades has focused on heuristics and biases affecting human judgments \citep{Tversky1974,Gigerenzer2008}. Given our limited cognitive capacities, we must often adapt to use approximations or "short-cuts" to find good-enough solutions to otherwise intractable problems. Heuristics are computationally cheap solutions that work well in environments characterized by the statistical properties of the environments they evolved in \citep{Tversky1983}. However, they can lead to systematic errors when these assumptions are violated. In the current work, the agreement-in-confidence heuristic provides a solution to the intractable problem of estimating the reliability of information in the absence of an objective standard. This solution works well when agreement correlates with accuracy, as is the case where initial judgment and advice are independent (Experiment 1). Thus, in Experiment 1, trust and influence patterns were little affected by the presence or absence of an objective standard. However, the process goes astray when the independence between judgment and advice is broken, as in Experiments 2 and 3 where advice was contingent on participants' initial decision and expressed confidence. Adverse consequences emerged in terms of distinct patterns of trust and influence with and without an objective standard. The result is explained by the fact that when feedback is removed participants are only left with their own initial probabilistic estimate. Any attempt to use this estimate to attribute feedback to their advisors is cursed by the error characterizing their own original judgment. As an example, imagine you are certain that you are drinking a very expensive wine at a blind wine tasting event. You will likely judge anybody who says otherwise as incompetent in wine matters. In the (not so unlikely) event that you are mistaken, you could keep thinking that a potentially brilliant sommelier is only a novice, with potentially negative consequences for you in terms of reputation and future wine choices. 

\paragraph{}
Our results were generally in line with the idea that we trust advisors according to their objective accuracy when feedback is present and agreeing-in-confidence advisors when it is absent. However, there were some notable (and surprising) exceptions. First, advisor agreement rate had an effect over and above accuracy even when feedback was available in Experiment 2. This finding suggests that multiple strategies might idiosyncratically converge to an estimate of advice reliability, with people relying on sometimes-useful cues (like agreement) even if they are strictly redundant in the current context. Second, people seem to value information over pure accuracy, as shown by participants in the Feedback group of Experiment 3, who preferred an advisor who agreed with them more frequently when they themselves were correct but low in confidence (the anti-bias advisor) over equally accurate but less informative advisors. This result is in agreement with literature on judge-advisor systems showing people's preference for advisors with unshared information \citep{VanSwol2007}. Finally, a result with a less clear explanation is that participants in Experiment 3 did not prefer the neutral to anti-bias advisor when feedback was absent. If anything, they tended to prefer the latter. We speculate that this unexpected result can be explained in light of the fact that, in domains where feedback is removed, expertise can still be estimated by comparing the classification variability observed for similar stimuli with the classification variability observed across different stimuli \citep{Weiss2003}. One possibility is that participants noticed the lack of variability in agreement rates (across different trial difficulties as experienced subjectively and expressed in terms of decision confidence) of the neutral advisor and consequently discounted this advisor.

\paragraph{}
When scaling up these cognitive mechanisms to larger population interactions as simulated in agent-based models, we observed that when feedback was not available and agents' judgments were independent, agreement-based trust formation strategies helped agents trust more accurate advisors. However, when individuals covaried in their signals (e.g., by sharing systematic biases), we observed the emergence of segregation of individuals according to their initial biases. Agents within a homogeneous population were more likely to trust and influence each other than agents belonging to different populations. The polarization of each cluster's average bias was reduced by the presence of random noise, increased signal strength or the presence of objective feedback. The results provide a potential new understanding of echo-chamber phenomena, already described in the literature \citep{Sunstein2001}, as a by-product of an otherwise adaptive mechanism - use of an agreement-in-confidence heuristic to estimate advice reliability - when it is difficult to objectively assess the reliability of others' opinions, and when these opinions might themselves be subject to systematic shared biases. In these scenarios, random fluctuations in initial bias for one or the other option within a population of equally accurate individuals can spiral out to form densely connected communities, thus effectively modifying the network structure. The results show how simulated agents endowed with realistic cognitive mechanisms can shed light on the emergence of complex patterns at the population level \citep{Epstein2013}. More broadly, the present research suggests the value of identifying strategies used by individual decision makers - who are likely to rely on imperfect heuristics given necessary limits in the information they access and cognitive resources available to process that information - and exploring how these strategies can influence behavior at the group and network level. Future research might usefully explore this approach further, and also the converse case of understanding how group and network dynamics might inform or constrain the decision strategies used by individuals.

\section{Conclusions}
The current work aimed to show that confidence is a valuable attribute of someone's judgment in social decision-making. It helps others discriminate when one is more likely to be correct (and thus value their contribution) but also helps a decision-maker to make consistent judgments about others irrespective of feedback availability. However, this potentially adaptive solution to an intractable problem of learning in the absence of feedback can backfire when judgments from different observers are not independent, leading to systematic biases in trust and influence that may be evident in the behavior of individuals and in the social networks they inhabit.  

\bibliography{main}
\bibliographystyle{apacite}

\end{document}


\maketitle

\section{Advice Value as Information Gain}
\subsection{Experiment 1}
We formalised an advisor's informational value as the mean absolute information gained after each possible social encounter with a specific advisor. Information gain is the difference between the posterior and prior probability of participant's correct response:

\begin{equation} \label{eq:2.1}
IG = p(d=w|e) - p(d=w)
\end{equation} 

where posterior probability correct $p(d=w|e)$ represents the probability that the decision $d$ is equal to the correct decision $w$, conditional on the specific social encounter $e$. Social encounter $e$ represents one of the four possible events: the advisor (1) confidently disagrees, (2) unconfidently disagrees, (3) unconfidently agrees, (4) confidently agrees (where "confidently" and "unconfidently" refer to the level of confidence expressed by the advisor on that trial). Posterior probability $p(d=w|e)$ was computed with Bayes' theorem and was proportional to participant's prior probability correct $p(d=w)$ and the likelihood of the social event given participant's accuracy $p(e|d=w)$. Given the staircase procedure, we used 70\% as prior $p(d=w)$. The probability of agreement (or disagreement) conditional on correct response and the overall probability of agreement (or disagreement) were known by design. The mean absolute information gain so computed (reported in the main text, Table 1) was lowest for the Inaccurate Uncalibrated advisor (0.08), intermediate for the Accurate Calibrated and Accurate Uncalibrated advisors (0.29 and 0.26 respectively) and the highest for the Calibrated but Inaccurate advisor (0.38). This can be intuitively understood by looking at Table 1, in the main text.  Although the Inaccurate Calibrated advisor's accuracy rate is lower than the Accurate Calibrated advisor, outcomes can be better predicted by its judgment. In particular, its judgments correlate strongly positively when sure and strongly negatively when unsure with the correct answer. On the contrary when the Accurate Calibrated advisor is unsure there is a much higher uncertainty about the final outcome. We also computed an expected information gain $IG_e$ for each advisor (Table 1 in the main text) by scaling $IG$ by the overall probability of each event:

\begin{equation} \label{eq:2.2}
IG_e = IG * p(e)
\end{equation} 

where $p(e)$ is the overall probability of each social event (i.e., confident disagreement, unconfident disagreement, unconfident agreement, confident agreement). The expected information gain captures the idea that extremely informative but very unlikely events are not very valuable. $IG_e$ values for each advisor were: Accurate Calibrated = .063, Accurate Uncalibrated = .063, Inaccurate Calibrated = .084, Inaccurate Uncalibrated = .021; suggesting that the Inaccurate Calibrated advisor's advice was the most informative.
\subsection{Experiment 2}
Similar to Experiment 1, we used conditional probabilities and the participants' expected accuracy to compute the informational value of each advisor. Advisors' mean absolute information gain $IG$ and expected information gain $IG_e$ were computed as in the previous experiment. Contrary to Experiment 1, however, advisors did not express different levels of confidence. This created only two possible social situations $e$ on each trial (instead of four as in the previous experiment), namely either agreement or disagreement. As can be seen in Table 2 in the main text, information gain was highest for the accurate advisors and the lowest for the agreeing but inaccurate advisor. 

\subsection{Experiment 3}
\begin{figure}[H]
\centering
  \includegraphics[width=1\textwidth]%
    {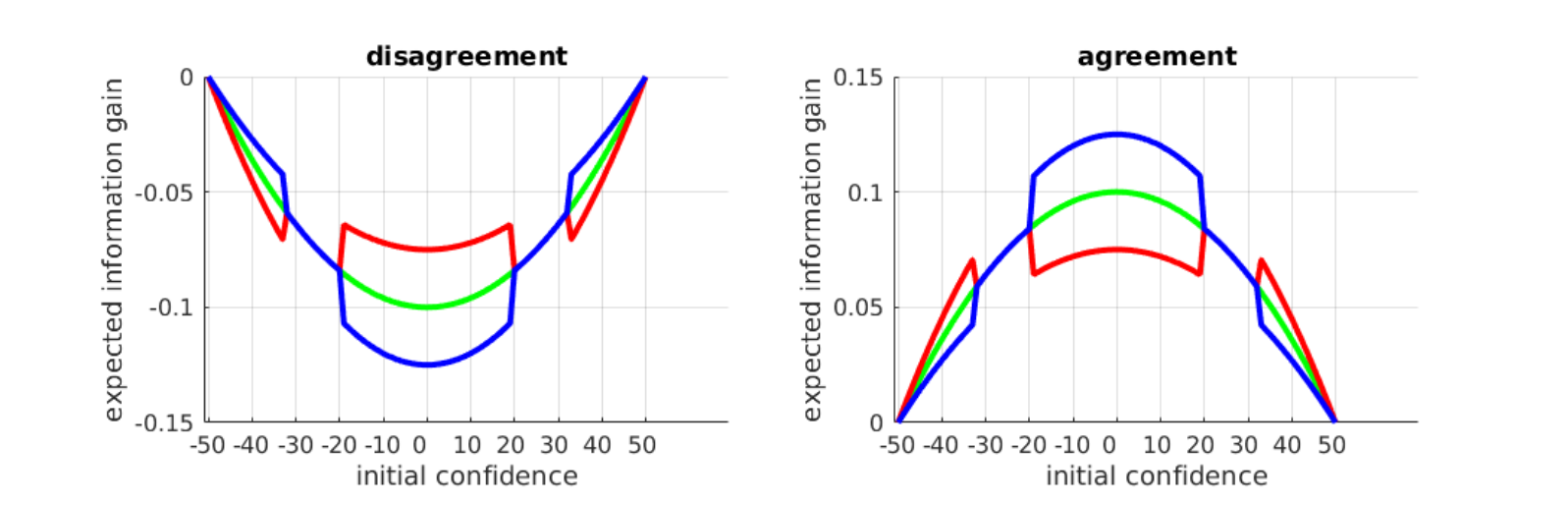}
  \caption[Experiment 3 - Information gain curves.]{Information gained after agreeing and disagreeing with each advisor type (color code) for each initial subjective prior confidence. Information gain is scaled by the likelihood of agreement and disagreement events. Advice informativeness can be quantified by the area under the curve.}
  \label{fig:exp1_IG_SI}
\end{figure}

The intuition that the anti-bias advisor was the most informative of the three advisors designed for Experiment 3 was confirmed using a numerical simulation based on an ideal Bayesian observer performing the task, with a Gaussian distribution of confidence centred on 25 and with a standard deviation of 10. For each initial confidence judgment the information gained from observing agreement or disagreement was computed for each advisor as the difference between posterior confidence and prior confidence. Contrary to previous Experiments, prior probability correct was here defined on a trial level based on pre-advice confidence. An expected information gain was computed by multiplying the information gain so obtained by the normalisation term in the Bayes formula. This produced a curve of expected information gains after agreeing or disagreeing with each advisor over possible pre-advice confidence levels (Figure \ref{fig:exp1_IG_SI}). The average area under the expected information gain curve was taken as an objective measure of advisor informativeness. Average areas under the curve were 14.74 for the unbiased advisor, 14.63 for the bias-sharing advisor and 15.78 for the anti-bias advisor. This procedure thus quantified and confirmed the intuition that the anti-bias advisor provided the most informative advice.

\section{Measures of interest}
Two measures of estimated advice reliability were defined. The first was the explicit trust that participants expressed in the advisors as collected by the brief questionnaires presented to participants every two blocks. In Experiment 1, four questions asked participants to directly rate on a scale from 1 ("Not at all") to 50 ("Extremely") how much they thought each advisor was accurate (Q1), confident (Q2a), trustworthy (Q3) and influential on their own choices (Q4). In Experiment 2, due to the absence of a confidence judgment from advisors, question 2 was replaced with a question asking about how much participants liked each advisor (Q2b: likeability question). For all Experiments, the first questionnaire was presented immediately after the practice blocks but before any interaction with the advisors took place so to provide a baseline measure. Baseline ratings were subtracted from following ratings to account for confounding factors related to advisors' appearance and inter-individual differences in the use of the scale. A principal component analysis (PCA) was performed for dimensionality reduction on normalised difference scores and the first component was taken as a unitary measure of expressed trust. 

In Experiment 1, question loadings for the Feedback condition were 0.52 (Q1), 0.44 (Q2a), 0.50 (Q3), 0.52 (Q4); and for the No-Feedback condition were 0.51 (Q1), 0.39 (Q2), 0.54 (Q3), 0.52 (Q4). 

In Experiment 2, question loadings for the Feedback condition were 0.53 (Q1), 0.39 (Q2b), 0.53 (Q3), 0.52 (Q4); and for the No-Feedback condition were 0.51 (Q1), 0.41 (Q2), 0.53 (Q3), 0.51 (Q4). 

In Experiment 3, question loadings for the Feedback condition were 0.53 (Q1), 0.42 (Q2b), 0.53 (Q3), 0.50 (Q4); and for the No-Feedback condition were 0.48 (Q1), 0.47 (Q2), 0.53 (Q3), 0.50 (Q4).

\paragraph{}
The second measure of interest was an implicit index of advisor's influence on participant's opinions, quantifying participant's confidence change from pre- to post-advice: 

\begin{equation} \label{eq:2.3}
\delta_{C}=C_{post} - C_{pre}. 
\end{equation} 

where $C_{pre}$ is, for Experiment 1 and 2, an integer value between +1 and +5 and $C_{post}$ is an integer value between -5 and +5 (negative $C_{post}$ representing changes of mind). Positive $\delta_{C}$ values mean increases in confidence from pre- to post-advice and negative values represent decreases in confidence. Notice that $\delta_{C}$ values have a negative skew, ranging from -10 (moving from highest confidence in one judgment to highest confidence in the opposite judgment) to +4 (moving from lowest to highest confidence rating for a single judgment). In Experiment 3, given the difference scale used, $C_{post}$ can assume values between -50 and 50, while $\delta_C$ can range from -100 to 49. Agreement and disagreement trials have typically opposite effects on confidence change: agreement usually leads to increases in confidence while disagreement to confidence decreases. The absolute magnitude of confidence shifts in both agreement and disagreement trials can be expected to grow larger as the participant makes more use of the advice received. Thus a unitary measure of influence was obtained by subtracting average $\delta_{C}$ in disagreement from average $\delta_{C}$ in agreement:

\begin{equation} \label{eq:2.4}
I=\bar{\delta}_{C}^a - \bar{\delta}_{C}^d 
\end{equation} 

where $I$ assumes greater values as participant's confidence increases in agreement and confidence decreases in disagreement become larger.

\section{Model Description}
Experiment 1 showed that people are sensitive to similar dimensions in the advice they receive both when objective feedback is available and when it was not provided. It is unclear what cues people are following to estimate their partners' reliability when feedback is taken away. Two simple explanations can be offered. The first one is that different advisors agreed differently often with participants and this in turn was taken as an indicator of good performance. In a binary choice task, if we assume that two people's judgments are independent, agreement rate between the two will scale linearly with the accuracy of each individual as long as performance is above chance. Given that participants knew their performance was greater than 50\% because they received a summary feedback at the end of each block, they might have used raw agreement with their partner as an indicator of the partner's good performance. Thus accumulating the number of agreement events over time for each individual advisor separately allows a subject to form a stable opinion about the other person's underlying accuracy. 

An alternative strategy participants could have used is to accumulate over time the estimated probability of the advice being correct on a given trial. This quantity could be generated based on the internal metacognitive signals of confidence, which is a noisy representation of the uncertainty associated with a given perceptual judgment. In other words, given that confidence in a decision is a probabilistic representation of the correctness of one's own decision, it can also be used to estimate the likelihood that the advice received is correct or incorrect. Accumulating such evidence over time can help a decision maker to estimate the reliability underlying advice whenever more secure signals are not available. 

To formalise such hypotheses we implemented a simple model that uses different pieces of information depending on different experimental conditions to estimate advisors' reliability. This simple model can then be compared with human observers to provide insight into the strategies they are using to evaluate advice reliability. 

Three different model variants are described below that account for the Feedback condition, the No-Feedback condition without metacognitive insight and the No-Feedback condition with metacognitive insight respectively. 

\subsection{Accuracy Model}
When objective feedback is given to participants by the experimenter, the model can use it to infer the accuracy rate of its advisors. The accuracy of the advisor ($Acc=\{0,1\}$) is the same as the accuracy of the subject in agreement trials, while is opposite in disagreement. By counting correct and error rates for each advisor separately, the model obtains a trial-by-trial estimation of the advisor's accuracy rate, $\theta$, as the ratio between the number of advisor's correct trials and the total encounters with that advisor:

\begin{equation}\label{theta}
\theta^i = \frac{\alpha^i}{\alpha^i + \beta^i}
\end{equation}

where $\alpha^i$ and $\beta^i$ are the correct and error counts respectively, during the past trials with advisor $i$:

\begin{equation} \label{positive}
\alpha^i = \sum\limits_{t=1}^n Acc_t
\end{equation} 

\begin{equation}\label{negative}
\beta^i = \sum\limits_{t=1}^n 1-Acc_t
\end{equation}

$t=1$ here represents the first encounter with advisor $i$ while $t=n$ represents the last one.  
A slight complication in Experiment 1, however, is that advisors also provided a binary confidence judgment associated with the advice. A simple way for the model to make use of advisor's confidence is by treating it as a linear scaler of the advice received. We applied a set of arbitrary weights to the four possible advice scenarios, namely the advisor is (1) correct and confident, (2) correct but unsure, (3) incorrect and unsure and (4) incorrect but confident (Table \ref{table:table3.1}). Although arbitrary, any set of weights that preserves the order of such events would result in similar final advisor preferences. 

\begin{table}[H]
\begin{tabular*}{\textwidth}{@{\extracolsep{\fill}}lcccc@{}} \toprule 
\multicolumn{1}{c}{} & \multicolumn{4}{c}{Event Observed} \\ \cmidrule(r){2-5}
 & Inaccurate & Inaccurate & Accurate & Accurate \\
 & Confident & Unsure & Unsure & Confident \\
Feedback &  -1 & -0.5 & +0.5 & +1 \\ \midrule
 & Disgree & Disagree & Agree & Agree \\
 & Confident & Unsure & Unsure & Confident \\
No-Feedback &  -1 & -0.5 & +0.5 & +1 \\ \bottomrule
\end{tabular*}
\caption[]{Model weights ($w$) applied to different advice events observed in the Feedback and No-Feedback scenario.}
\label{table:table3.1}
\end{table}

Thus instead of simple accuracies, $\alpha$ and $\beta$ in equations \ref{positive} and \ref{negative} can now be reformulated as:

\begin{equation} \label{positivew}
\alpha^i = \sum\limits_{t=1}^n .5+.5*w_t
\end{equation} 

\begin{equation}\label{negativew}
\beta^i = \sum\limits_{t=1}^n .5-.5*w_t
\end{equation}

This set of equations results in values of 1, 0.75, 0.25 and 0 for the four events listed above respectively. Although these values could be simply summed to obtain $\alpha$ and $\beta$ values, the unusual formulation of the equations \ref{positivew} and \ref{negativew} was preferred to be coherent with the equations describing the following models. They show how a simple model can take into account feedback, advice received and advisors' expressed confidence to track over time the objective reliability of its advisors.

\subsection{Consensus Model}
When feedback is removed from the participants, as in the No-Feedback condition of Experiment 1, the model does not have access to the advisors' objective accuracy. It must then rely on different proxies for objective accuracy and integrate those instead over time. The first cue to underlying accuracy rate we considered is agreement rate. When two independent agents express judgments on a binary task, the agreement rate between the two linearly scales with the accuracy of each whenever the accuracy rate is higher than chance: $Agr = Acc_1 * Acc_2 + (1-Acc_1)*(1-Acc_2)$. We thus adapted the equations of the \textit{Accuracy} model above to exploit this covariation.
Instead of tracking the accuracy rates of its advisors, the \textit{Consensus} model tracks their agreement rates with subjective judgments. Thus equations \ref{positivew} and \ref{negativew} can be used to estimate a $\theta$ value by now using as $w_t$ the scaled agreement observed on encounter $t$ as described in Table \ref{table:table3.1}. To take into account the fact that in Experiment 1 advisors expressed a binary confidence judgment themselves associated with the advice, we used the same linear weights applied to the \textit{Accuracy} model also to scale agreement (Table \ref{table:table3.1}). In other words this model perfectly conflates accuracy with agreement, assuming that whenever an advisor agrees with the subjective original judgment, the advisor must be correct. Although this clearly is a simplifying assumption, the model offers a useful proof of concept to understand what inferences an agent lacking metacognitive insight can make simply by using heuristics. It can thus provide a benchmark to quantify the information that is present in the advice received.

\subsection{Confidence Model}
A more nuanced strategy that could be employed to estimate advisors' reliability when feedback is not directly available is through use of internal metacognitive signals. Trial-level variability in subjective confidence is known to covary with objective accuracy in a perceptual task \citep{Henmon1911} and it theoretically represents the estimated likelihood of having made a correct judgment and/or selected the correct response \citep{Pouget2016}. Thus, instead of simply using agreement rates as a cue for accuracy rate, a model endowed with metacognitive insight could accumulate over time the subjective probability that an advisor expressed a correct judgment. A \textit{Confidence} model was created under the assumption that the trial-by-trial subjective reports of confidence are directly related to the true underlying estimated probabilities of having chosen the correct answer. On agreement trials the model estimates the probability of the advice being correct as the subjective probability of a correct answer. Conversely on disagreement trials the model estimates the probability of the advice being correct as the probability of having itself made an error. In other words trial-level agreement ($Agr=\{0,1\}$) is scaled by trial-confidence expressed as a probability over outcomes (correct vs. incorrect response). Thus equations \ref{positivew} and \ref{negativew} above become according to this model:

\begin{equation} \label{pos_confm}
\alpha^i = \sum\limits_{t=1}^n .5+(p_t(corr)-.5)*w_t
\end{equation} 

\begin{equation}\label{neg_confm}
\beta^i = \sum\limits_{t=1}^n .5-(p_t(corr)-.5)*w_t
\end{equation}

where $w_t$ represents the scaled trial-level agreement as described in Table \ref{table:table3.1} and $p(corr)$ represents pre-advice confidence. As described below, rather than taking participants' confidence as a pure index of subjective $p(corr)$, we transformed the value to (1) reduce inter-subjects variability and (2) increase scale sensitivity. Regardless, the crucial point is that this model capitalises on the fact that being in agreement or disagreement with an advisor is more informative when the model is itself confident that it gave a correct answer than when it is more likely to have made a mistake.

\paragraph{Experiments 2-3.}
In Experiments 2 and 3, advisors did not express a level of confidence with their judgments. This allowed to simplify the above equations describing the three models. The \textit{Accuracy} model could be simplified using equations \ref{positive} and \ref{negative} to compute $\alpha$ and $\beta$ for each advisor instead of equations \ref{positivew} and \ref{negativew}. Similarly, the \textit{Consensus} model now computes $\alpha$ and $\beta$ values for each advisor $i$ separately as:

\begin{equation} \label{positive_agr}
\alpha_i = \sum\limits_{t=1}^n .5+.5*Agr_t
\end{equation} 

\begin{equation}\label{negative_agr}
\beta_i = \sum\limits_{t=1}^n .5-.5*Agr_t
\end{equation}

where $Agr_t$ is the partner's consensus ($Agr=\{-1,1\}$) observed on encounter $t$.
Finally, the simplified \textit{Confidence} model computes $\alpha$ and $\beta$ values as:

\begin{equation} \label{positive_conf}
\alpha = \sum\limits_{t=1}^n .5+(p(corr)-.5)*Agr_t
\end{equation} 

\begin{equation}\label{negative_conf}
\beta = \sum\limits_{t=1}^n .5-(p(corr)-.5)*Agr_t
\end{equation}

where $p(corr)$ is the pre-advice confidence expressed in probability scale as described in equation \ref{conf2proba}.

\subsection{Bayesian update}
All model variants can use the current estimated advisor's reliability $\theta$ to appropriately update the pre-advice probability of having selected the correct answer $p(corr)$ into a normative posterior, based on the binary advice $A$ received (agree vs. disagree): 

\begin{equation} \label{bayes}
p(corr|A^i) = \frac{p(corr)p(A^i|corr)}{p(corr)p(A^i|corr)+p(err)p(A^i|err)}
\end{equation}

where $p(err)$ is the subjective probability of making a mistake on the current trial and $p(A^i|corr)$ is the probability that advisor $i$ agrees or disagrees given that the participant's choice is correct. Prior probability $p(corr)$ is estimated from a simple linear transformation of the pre-advice trial-level confidence data obtained from the participants after appropriate pre-processing. Pre-processing consisted in a parameter-free transformation that (a) brings all subjective confidence distributions on to a similar scale thus reducing the inter-subject variability and (b) expands the centre of the original subjective confidence distributions so to increase the informativeness of the average trial. This operation was inspired by recent models of adaptive information gain control \citep{Cheadle2014}. According to these proposals, the brain adapts the gain of neuronal firing to the range of information available over different time scales and cognitive domains \citep{Carandini2011,Cheadle2014}. Here it serves the purpose of increasing the discriminability or information gain of different trials so that trials that are close together on confidence scale gets pulled apart on to a probability scale. The transformation uses parameters obtained from the data:

\begin{equation}
\label{eq:preprocessing}
\hat{C}_{pre} = N * normcdf(C_{pre})
\end{equation}

where $normcdf(C)$ is the normal cumulative density function of the pre-advice confidence $C$ ratings distribution, and $N$ is the number of confidence ratings available on each interval of the scale (in Experiments 1,2: N = 5; in Experiment 3: N = 50). This simple transformation has the property of translating a normal distribution into a uniform distribution in the range $[0, N]$. Notice that this transformation does not affect the ranking of confidence judgments but only their spacing along a probability scale. After pre-processing, confidence ratings were translated into a probability scale with the linear transformation:

\begin{equation} \label{conf2proba}
p(corr) = 0.5+(0.1-\epsilon)*\hat{C}_{pre}
\end{equation}

where $\epsilon$ is a small jitter ($\epsilon=.002$) introduced to avoid maximum confidence ratings being turned into probability of one and zero, which would in turn cause inconsistencies within the Bayesian formula (e.g., no confidence change regardless of advice reliability). Thus $p(corr)$ represents trial-level confidence on a probability scale, which can be interpreted as the probability that the participant assigns to having given a correct answer on a given trial. From $p(corr)$ we can also derive the subjective probability that a given trial will end up in an error: $p(err) = 1-p(corr)$.

To estimate the likelihood term $p(A^i|corr)$ in equation \ref{bayes} we applied a simple heuristic that uses the reliability $\theta$ of a given advisor:

\begin{equation} \label{likelihood}
p(A^i|corr) = \theta^A * (1-\theta)^{1-A}
\end{equation}

The equation above simply states that the probability of observing advisor $i$'s agreement ($A^i=1$) when the participant is correct is equal to the accuracy rate of the advisor itself, assuming advisor's and participant's judgments are independent. Instead the probability of observing disagreement ($A^i=0$) on the same trials is the advisor's error rate. In other words, the probability of agreement in trials when the participant is correct is the probability that the advisor too is correct. Similarly the probability of disagreement in trials when the participant is correct is equal to the probability that the advisor is wrong. 

\subsection{ABM results}

Discounting advice was obtained by a linear regression toward the uncertainty point .5 using the following equation:
\begin{equation}
    E_s' = 0.5 + ( \theta * (p'(A) - 0.5))
    \label{eq:discounting}
\end{equation}

The following figures supplement, figures in the main text.

\begin{figure}[H]
\centering
  \includegraphics[width=\textwidth]%
    {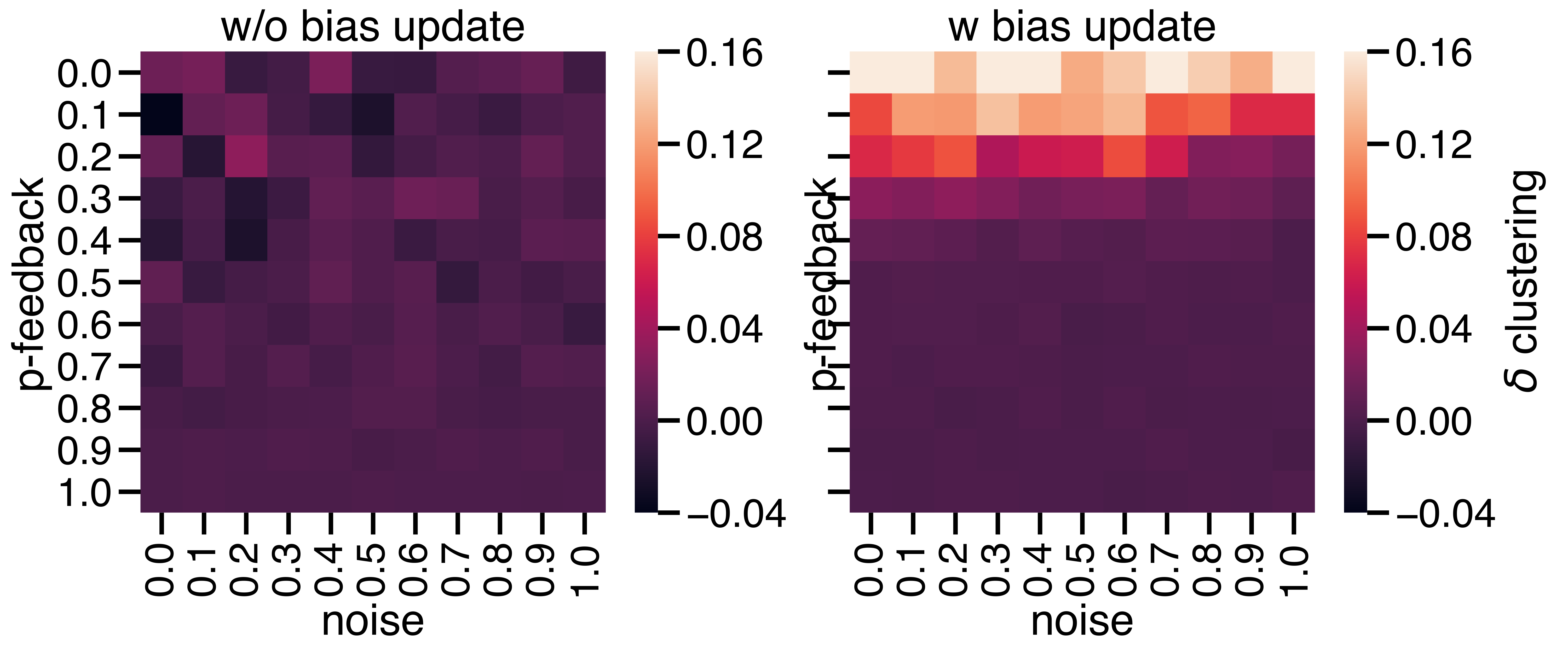}
  \caption[Agent-based model - Results]{Clustering difference between agents who use discounting and agents who don't. Left panel: agents do not update their bias. Right panel: agents update their bias.}
  \label{fig:ABMclustering}
\end{figure} 

\begin{figure}[H]
\centering
  \includegraphics[width=\textwidth]%
    {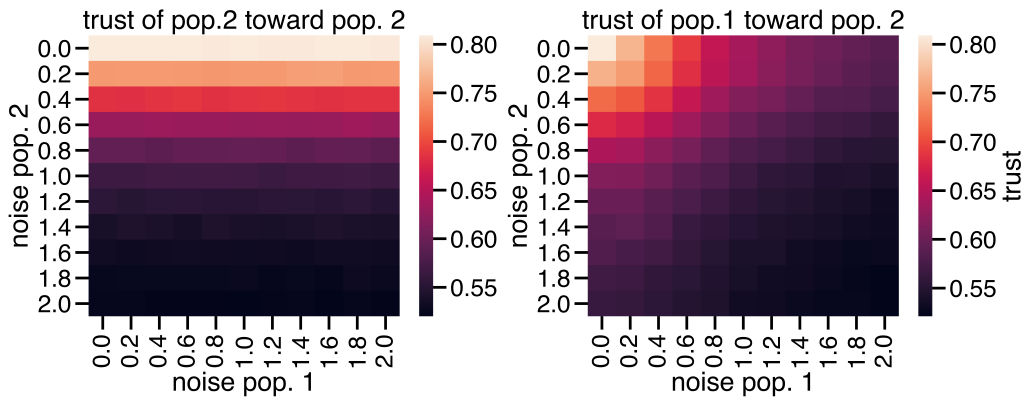}
  \caption[Agent-based model - Results]{Trust of each subpopulation toward population 2.}
  \label{fig:ABMtrust_split}
\end{figure} 

\begin{figure}[H]
\centering
  \includegraphics[width=\textwidth]%
    {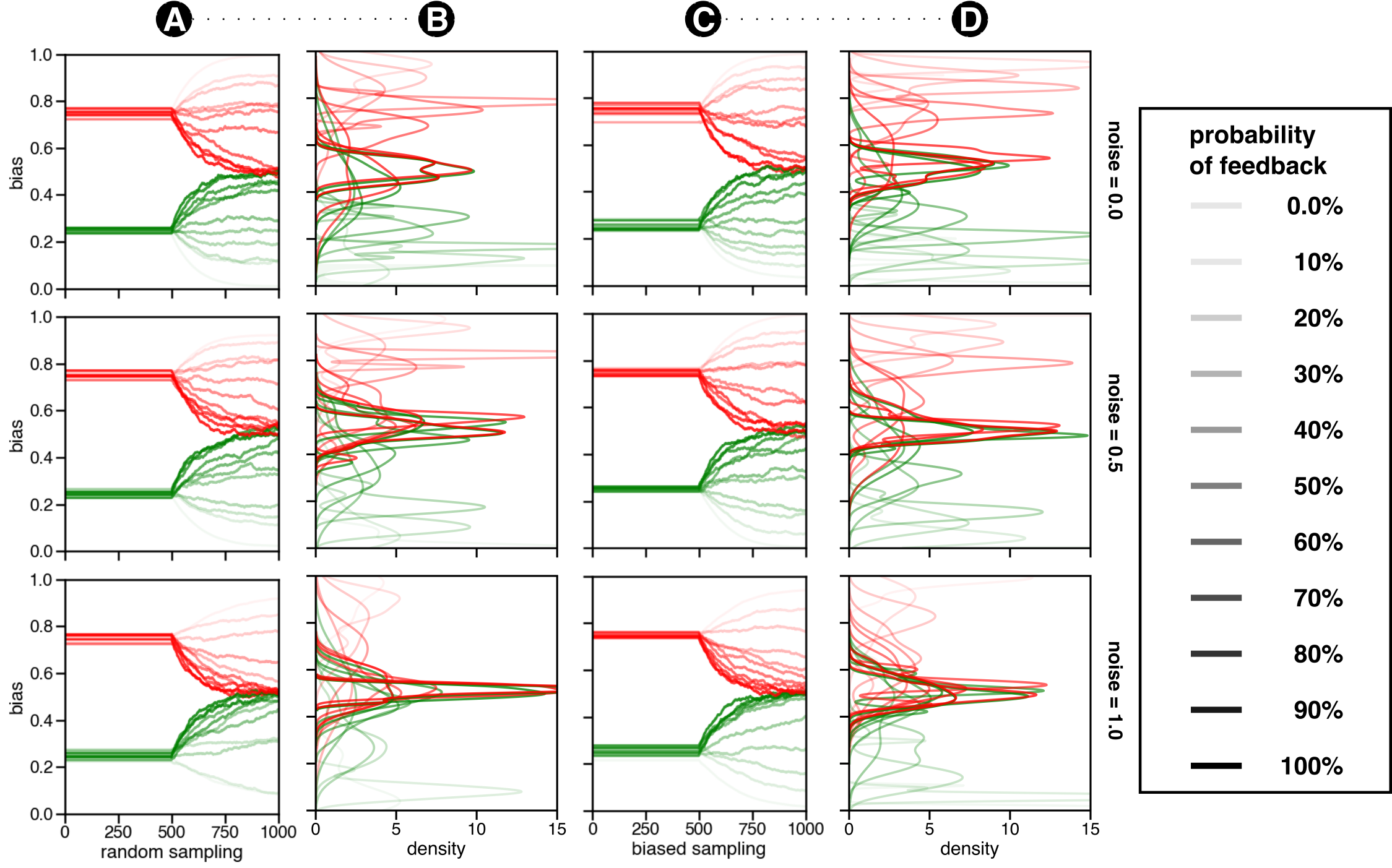}
  \caption[Agent-based model - Results]{Bias distribution and evolution in simulations where agents discount advice proportionally to trust in the advisor.}
  \label{fig:ABMbias}
\end{figure} 

\begin{figure}[H]
\centering
  \includegraphics[width=\textwidth]%
    {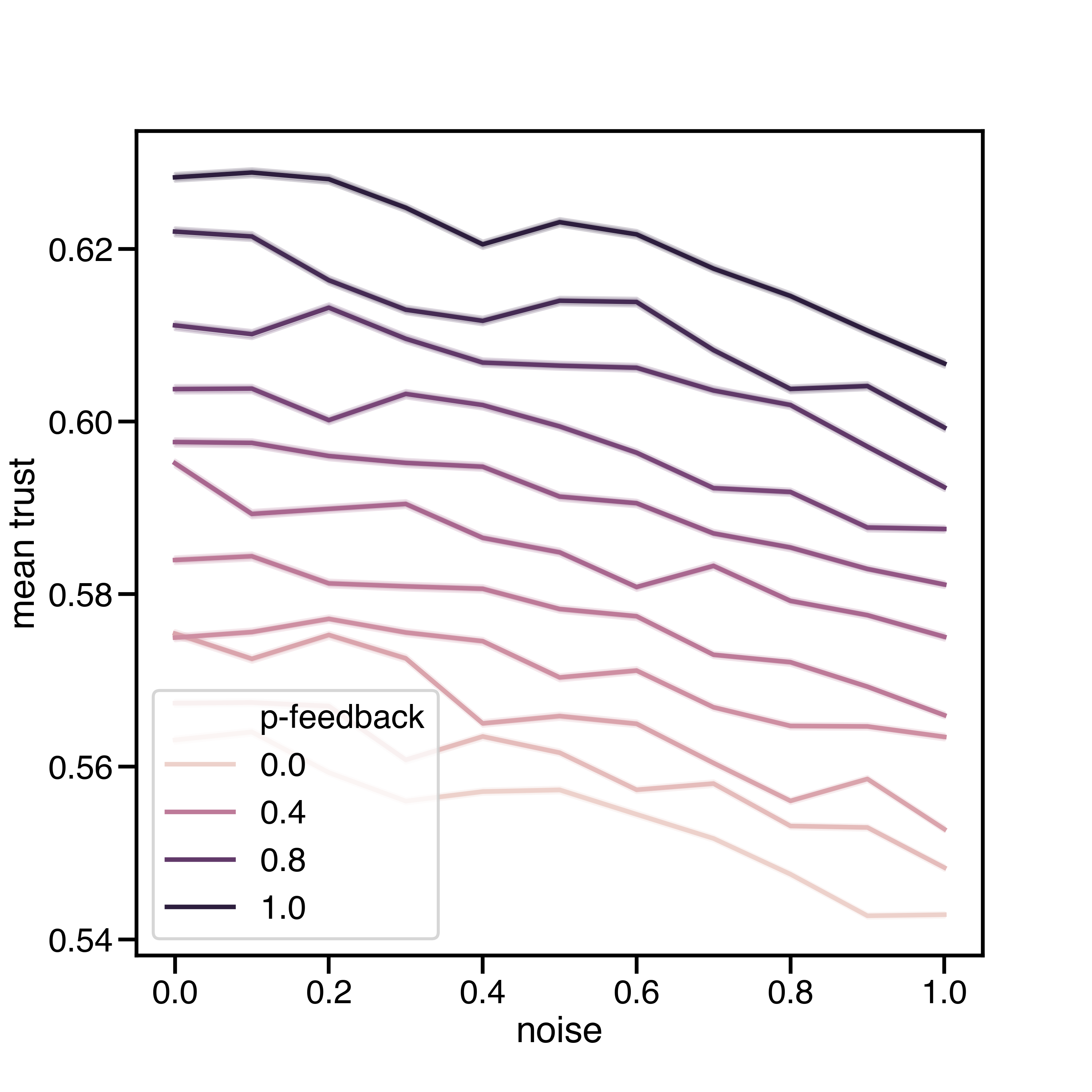}
  \caption[Agent-based model - Results]{Average trust as a function of probability of feedback and noise. Trust appears to decrease as noise increases and feedback availability decreases.}
  \label{fig:ABMtrust}
\end{figure} 

\begin{figure}[H]
\centering
  \includegraphics[width=\textwidth]%
    {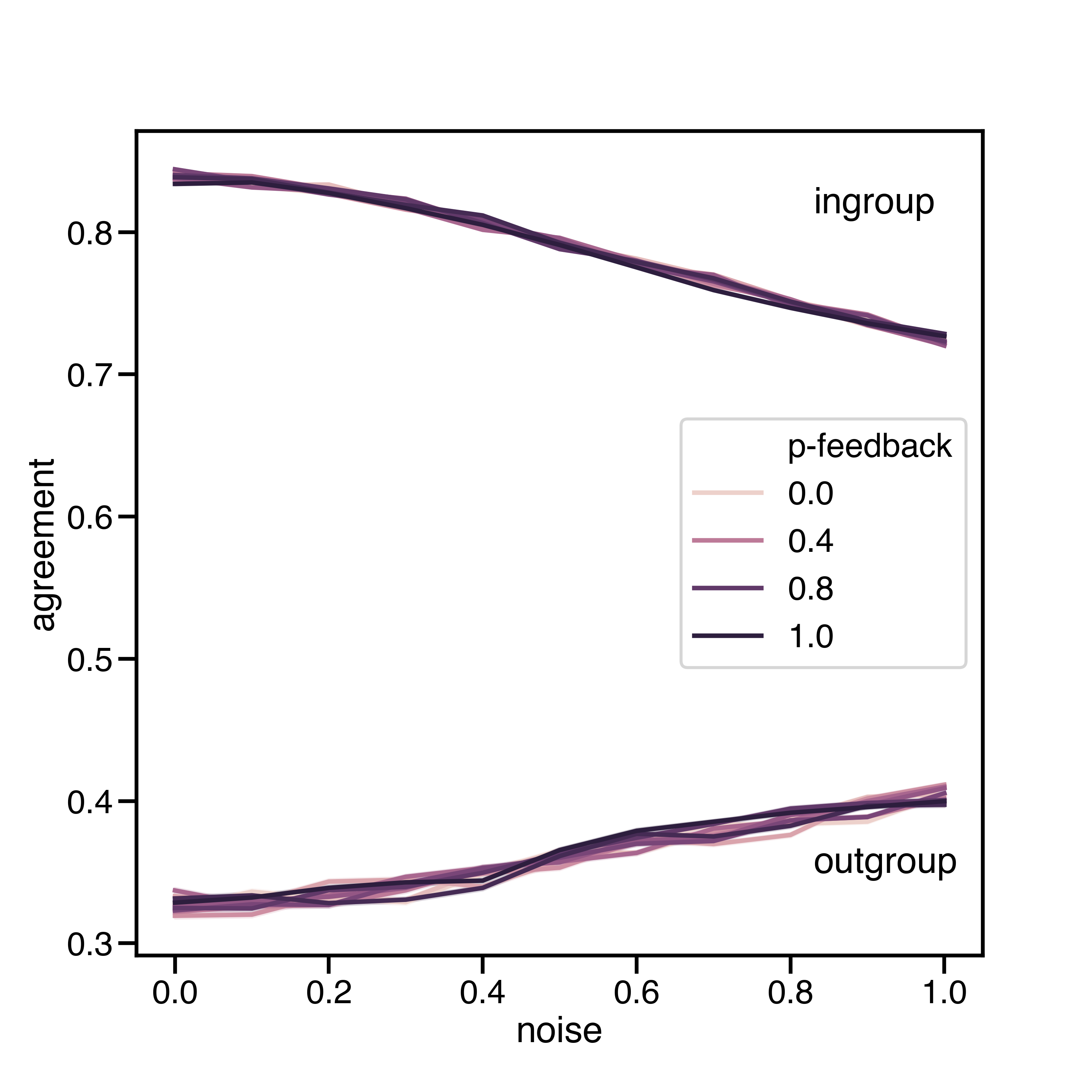}
  \caption[Agent-based model - Results]{Average agreement rate as a function of probability of feedback and noise. Agreement with ingroup appears to decrease as a function of increasing noise, while agreement with outgroup tends to increase as a function of noise. On the contrary, feedback availability does not affect agreement rates.}
  \label{fig:ABMagreement}
\end{figure} 

\bibliography{main}
\bibliographystyle{apacite}